\documentclass[fleqn,usenatbib]{mnras}
\usepackage{newtxtext,newtxmath}
\usepackage[T1]{fontenc}
\usepackage{array}

\usepackage{amsmath}
\usepackage{bm}
\usepackage{graphicx}
\usepackage{xcolor}
\usepackage{hyperref}
\usepackage{pdflscape}

\usepackage{multirow}

\DeclareRobustCommand{\VAN}[3]{#2}
\let\VANthebibliography\thebibliography
\def\thebibliography{\DeclareRobustCommand{\VAN}[3]{##3}\VANthebibliography}

\title[Pulsar scintillation through thick and thin]{Pulsar scintillation through thick and thin: Bow shocks, bubbles, and the broader interstellar medium}

\author[S.K. Ocker et al.]{Stella Koch Ocker,$^{1,2,3}$\thanks{E-mail: socker@caltech.edu}
James M. Cordes,$^{1}$
Shami Chatterjee,$^{1}$
Daniel R. Stinebring,$^{4}$
Timothy Dolch,$^{5,6}$\newauthor 
Vincent Pelgrims,$^{8}$
James W. McKee,$^{9,10}$
Christos Giannakopoulos,$^{5,7}$
Daniel J. Reardon$^{11,12}$
\\
$^{1}$Department of Astronomy and Cornell Center for Astrophysics and Planetary Science, Cornell University, Ithaca, NY 14850, USA\\
$^{2}$Cahill Center for Astronomy and Astrophysics, California Institute of Technology, Pasadena, CA 91101, USA\\
$^{3}$The Observatories of the Carnegie Institution for Science, Pasadena, CA 91101, USA\\
$^{4}$Department of Physics and Astronomy, Oberlin College, Oberlin, OH 44074, USA \\
$^{5}$Department of Physics, Hillsdale College, Hillsdale, MI 49242, USA\\
$^{6}$Eureka Scientific, 2452 Delmer Street, Suite 100, Oakland, CA 94602-3017, USA\\
$^{7}$Department of Physics, University of Cincinnati, Cincinnati, OH 45221, USA\\
$^{8}$Universit{\'e} Libre de Bruxelles, Science Faculty CP230, B-1050 Brussels, Belgium\\
$^{9}$E. A. Milne Centre for Astrophysics, University of Hull, Cottingham Road, Kingston-upon-Hull, HU6 7RX, UK \\
$^{10}$Centre of Excellence for Data Science, Artificial Intelligence and Modelling (DAIM), University of Hull, Cottingham Road, Kingston-upon-Hull, HU6 7RX, UK\\
$^{11}$Centre for Astrophysics and Supercomputing, Swinburne University of Technology, P.O. Box 218, Hawthorn, Victoria 3122, Australia\\
$^{12}$OzGrav: The Australian Research Council Centre of Excellence for Gravitational Wave Discovery, Hawthorn VIC 3122, Australia
}

\date{Accepted 2023 November 24. Received 2023 November 18; in original form 2023 September 25}

\pubyear{2023}

\begin{document}
\label{firstpage}
\pagerange{\pageref{firstpage}--\pageref{lastpage}}
\maketitle

\begin{abstract}
Observations of pulsar scintillation are among the few astrophysical probes of very small-scale ($\lesssim$ au) phenomena in the interstellar medium (ISM).
In particular, characterization of scintillation arcs, including their curvature and intensity distributions, can be related to interstellar turbulence and potentially over-pressurized plasma in local ISM inhomogeneities, such as supernova remnants, HII regions, and bow shocks.
Here we present a survey of eight pulsars conducted at the Five-hundred-meter Aperture Spherical Telescope (FAST), revealing a diverse range of scintillation arc characteristics at high sensitivity. These observations reveal more arcs than measured previously for our sample. At least nine arcs are observed toward B1929$+$10 at screen distances spanning $\sim 90\%$ of the pulsar's $361$ pc path-length to the observer. Four arcs are observed toward B0355$+$54, with one arc yielding a screen distance as close as $\sim10^5$ au ($<1$ pc) from either the pulsar or the observer. Several pulsars show highly truncated, low-curvature arcs that may be attributable to scattering near the pulsar. The scattering screen constraints are synthesized with continuum maps of the local ISM and other well-characterized pulsar scintillation arcs, yielding a three-dimensional view of the scattering media in context.
\end{abstract}


\section{Introduction}

Pulsars emit coherent radio beams that are scattered by electron density fluctuations along the line-of-sight (LOS), leading to wavefield interference at the observer. The resulting intensity modulations, or scintillation, are observed in pulsar dynamic spectra (intensity as a function of frequency and time) as highly chromatic and variable on minute to hour timescales. The character of the scintillation pattern (rms intensity modulations and timescale) is related to the Fresnel scale $\sim \sqrt{\lambda D}$, which is $\sim 0.01$ astronomical units (au) for typical pulsar distances $D$ and observing wavelengths $\lambda$ \citep{rickett1990}. Pulsar scintillation can thus probe $\lesssim$ au structure in the interstellar medium (ISM). Numerous studies have demonstrated that such structure not only exists but may be ubiquitous \citep{zweibel_review18}. 

Pulsar dynamic spectra often exhibit organized interlacing patterns, which manifest as parabolas of power in the secondary spectrum obtained by 2D Fourier Transform (FT) of the dynamic spectrum \citep{stinebring2001}. These scintillation arcs are widely observed \citep{stinebring2022,main2023}, and their parabolic form is understood to be a generic result of the square-law relationship between time delay and angle in small-angle scattering \citep{walker2004,cordes2006}. However, observed secondary spectra show a broad range of arc characteristics for different LOSs and for single LOS at different epochs and observing frequencies, including variable arc widths, inverted arclets, clumps and asymmetries in arc intensity, and multiple arcs of different curvature (for examples see Figures~\ref{fig:all_spec1}-\ref{fig:all_spec2}, along with \citealt{stinebring2022} and \citealt{main2023}). Efforts to connect these features to the underlying physics of the ISM have largely focused on two scenarios that need not be mutually exclusive \citep{cordes1986}: diffraction through a turbulent cascade of density fluctuations, and refraction through discrete structures not necessarily associated with ISM turbulence \citep[e.g.][]{pen2014}. We emphasize that both diffraction and refraction can produce scintillation arcs in their most generic forms.

One method to assess the plasma properties of the scintillating medium involves mapping the secondary spectrum to the pulsar scattered image. This method has been applied to both single-dish pulsar observations under limiting assumptions \citep[e.g.][]{stinebring2019,rickett2021,hengrui2022}, and to Very Long Baseline Interferometry (VLBI) observations yielding scattered images at much higher resolution, the most well-studied example being PSR B0834$+$06 \citep{brisken2010,liu2016,simard2019,baker2022b}. This pulsar exhibits reverse arclets and arc asymmetries \citep{hill2005}, with a reconstructed scattered image that is highly anisotropic and contains discrete inhomogeneities on sub-au scales \citep{brisken2010}. These features have been attributed to either highly over-pressurized plasma structures \citep{hill2005} or plasma ``sheets'' viewed at large inclination angles relative to the LOS \citep{pen2014,liu2016,simard2018}. 

Complementary constraints on the plasma responsible for scintillation arcs can be obtained from the secondary spectrum itself. The primary constraint of interest is the arc curvature, which can be used to infer the distance to the scattering medium and hence determine the spatial scale of the scattered image, which is related to the spatial scale of the plasma density fluctuations. Precise scattering screen distances are typically obtained by measuring temporal variations in arc curvature, which occur periodically due to the Earth's motion around the Sun and the pulsar's orbital motion around its companion, if it has one \citep[e.g.][]{main2020,reardon2020}. Additional constraints from the secondary spectrum include comparing the observed arc frequency dependence and arc power distributions to theoretical expectations for a turbulent medium, which have yielded mixed evidence for and against Kolmogorov turbulence among different LOSs \citep[][]{hill2003,stinebring2019,reardon2020,rickett2021,turner2023}. In some cases, arc power is observed to vary systematically over time, presumably due to discrete ISM structures crossing the LOS \citep{hill2005,wang2018,sprenger2022}. Scattering screens have been connected to foreground structures observed at other wavelengths, including an HII region \citep{mall2022}, local interstellar clouds \citep{mckee2022}, a pulsar supernova remnant \citep{yao2021}, and the Monogem Ring \citep{yao2022}. Scintillation arcs have also been associated with screens near the edge of the Local Bubble \citep[e.g.][]{bhat2016,reardon2020,mckee2022,yao2022,liu2023L}, but the connection between these arcs and Local Bubble properties remains unclear.

A less explored source of scintillation arcs is stellar bow shocks, including those of pulsars. While many pulsars have transverse velocities in excess of the fast magnetosonic speed expected in the ISM, only a handful of pulsar bow shocks have been observed through direct imaging of the forward shock and/or the ram pressure confined pulsar wind nebula (PWN; \citealt{kargaltsev2017}). 
Recently, scintillation arcs have been detected from the bow shock of a millisecond pulsar (D. Reardon et al., submitted). This result raises
the possibility of using scintillation arcs to probe pulsar bow shocks, independent of more traditional direct imaging techniques \citep[e.g.][]{brownsberger2014}.

We have used the Five-hundred-meter Aperture Spherical Telescope (FAST) to observe eight pulsars with flux densities, spin-down luminosities, and transverse velocities favorable for carrying out a survey of scintillation arcs from near-pulsar screens, including bow shocks (see Table~\ref{tab:source_list} and Section~\ref{sec:theory}). Scintillation arcs are faint features and are typically analyzed from data recorded with narrow ($\sim100$\,kHz) channel widths, leading to a very low per-channel signal-to-noise (S/N). Therefore the high gain of FAST allows for highly sensitive measurements. Our observations yielded a rich array of scintillation arcs probing a broad distribution of screens between the pulsars and observer. In this study we present the results of our FAST observing campaign, with an emphasis on constraining the roles of the extended ISM and discrete local structures, including pulsar bow shocks, in observed secondary spectra. Our sample is well-suited to address these issues as it spans a range of distances, dispersion measures (DMs), and scintillation regimes.

The paper is organized as follows: Section~\ref{sec:theory} presents the basic theory of scintillation arcs and requirements to detect near-pulsar screens; Section~\ref{sec:obs} describes the FAST observations; Section~\ref{sec:analysis} shows the data analysis techniques implemented; and results for each pulsar are given in Section~\ref{sec:results}. Section~\ref{sec:screen_constraints} discusses the connection between scintillation arc properties and the ISM, including turbulent density fluctuations, bow shocks, and known discrete structures. We present conclusions in Section~\ref{sec:conc}.

\section{Theory of Dynamic \& Secondary Spectra}\label{sec:theory}

\subsection{Basic Phenomenology}

The pulsar dynamic spectrum consists of fringe patterns that form from interference of scattered waves. We assume that the scattering occurs in a thin screen, and discuss the potential relevance of extended media in Section~\ref{sec:conc}. The interference for two scattering angles (or equivalently, two angular locations in the pulsar scattered image) $\pmb{\theta}_1$ and $\pmb{\theta}_2$ leads to a sinusoidal fringe that corresponds to a single Fourier component in the secondary spectrum at coordinates $f_t,f_\nu$ \citep{stinebring2001,walker2004,cordes2006}:\footnote{{Note the minus sign in Eq.~\ref{eq:ft}: $f_t < 0$ when $\theta_2 > 0$ and vice versa, due to the relative motion of the pulsar and the deflector. E.g., \cite{hill2005} observed individual arclets that moved from negative to positive $f_t$ over the course of several months, due to the pulsar LOS passing through a discrete scattering structure.}} 
\begin{align}
    f_t &= -\frac{1}{s\lambda} (\pmb{\theta}_2 - \pmb{\theta}_1) \cdot \pmb{V}_\perp \label{eq:ft} \\
    f_\nu &= \bigg[ \frac{D(1-s)}{2cs}\bigg]  (\pmb{\theta}_2^2 - \pmb{\theta}_1^2) \label{eq:fnu}.
\end{align}
Here $D$ is the distance between the pulsar and observer, $c$ is the speed of light, $\lambda$ is the observing wavelength at the center of the band, $s$ is the fractional screen distance ($s=0$ at the pulsar and $s=1$ at the observer), and $\pmb{V}_\perp$ is the transverse velocity of the screen where it intersects the direct LOS,
\begin{equation}\label{eq:vperp}
    \pmb{V}_\perp = (1-s)\pmb{V}_{\rm psr\perp} + s\pmb{V}_{\rm obs\perp} - \pmb{V}_{\rm scr\perp}.
\end{equation}
The secondary spectrum coordinates given by Eq.~\ref{eq:ft}-\ref{eq:fnu} are the Fourier conjugates of time $t$ and frequency $\nu$, and are equivalent to the differential Doppler shift and the differential Doppler delay. It can also be convenient to consider the Fourier conjugate to observing wavelength \citep{cordes2006,fallows2014,reardon2020}, 
\begin{equation}\label{eq:flam}
    f_\lambda = c f_\nu/\lambda^2 = \bigg[ \frac{D(1-s)}{2s\lambda^2}\bigg]  (\pmb{\theta}_2^2 - \pmb{\theta}_1^2).
\end{equation}
From Eq.~\ref{eq:ft}-\ref{eq:fnu} it is apparent that each interfering pair of points $(\pmb{\theta}_1,\pmb{\theta}_2)$ lies on a parabola $f_\nu \propto f_t^2$ due to the linear and quadratic relationships between $\pmb{\theta}$ and $f_t$ and $f_\nu$, respectively.  
A scintillation arc can result from interference between all angular pairs \citep{cordes2006}. 
\begin{landscape}
\centering
\begin{figure}
    \centering
    \includegraphics[width=1.25\textwidth]{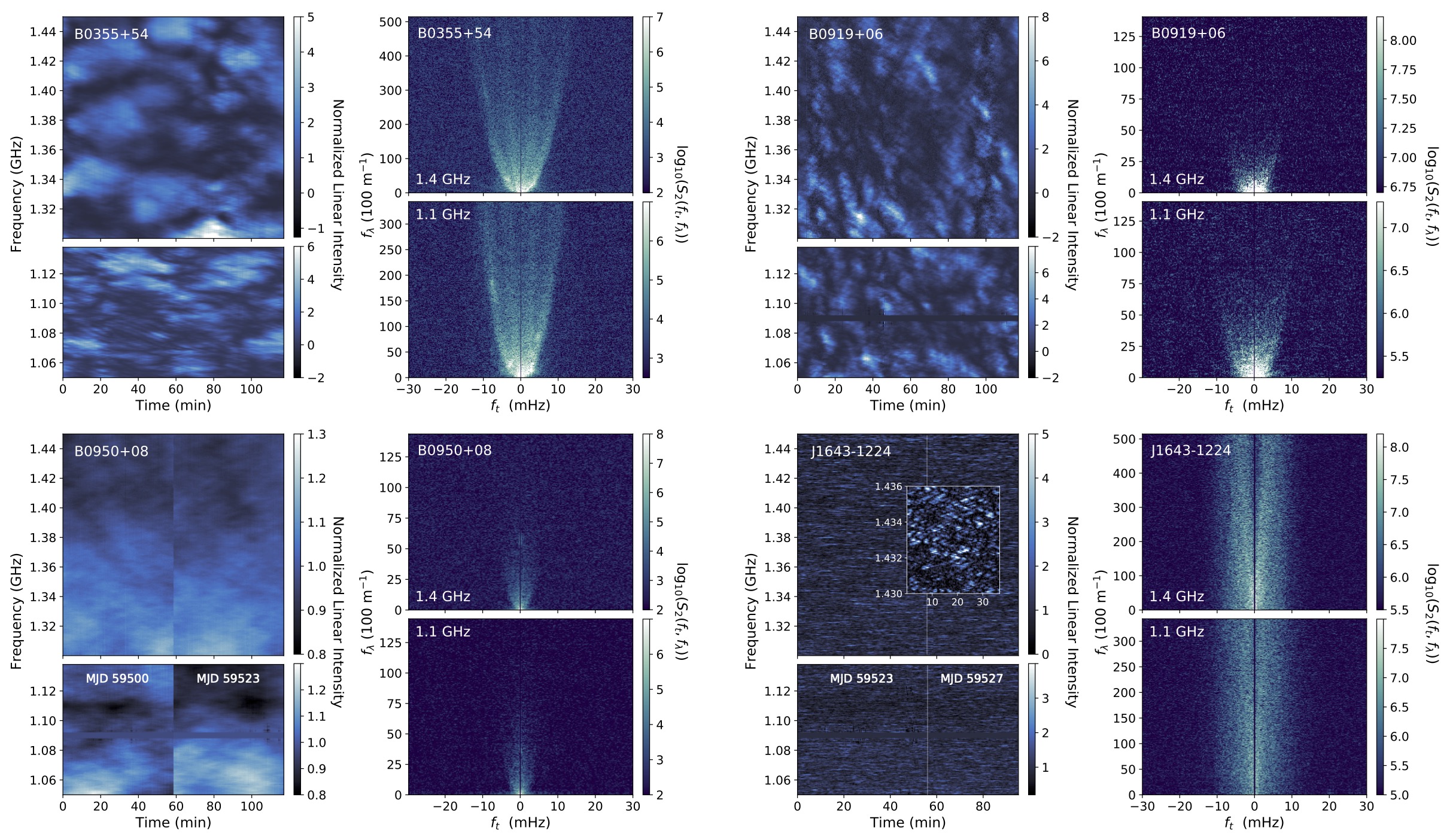}
    \caption{Dynamic and secondary spectra for (top left to bottom right): PSRs B0355$+$54, B0919$+$06, B0950$+$08, and J1643$-$1224. Each dynamic spectrum has a 3.2 s subintegration time and a 0.06 MHz frequency resolution, and has been normalized to unit mean linear intensity. Some dynamic spectra have strong radio frequency interference (RFI) at 1090 MHz that is masked and set to the mean intensity for visualization here, but to calculate secondary spectra the dynamic spectra were interpolated across the bad frequency channel using a 2D Gaussian kernel. Each secondary spectrum was calculated using the dynamic spectrum interpolated onto an equispaced wavelength grid, and is shown here in logarithmic power units. The mean (log$_{10}$) off-arc noise baseline has been subtracted in the secondary spectra shown, and the $f_t=0$ channel masked. For B0950$+$08, secondary spectra for the second epoch (MJD 59523) are shown. For J1643$-$1224, an inset shows a 6 MHz and 35 min-long portion of the dynamic spectrum on MJD 59527. The secondary spectra shown for J1643$-$1224 correspond to the first, longer epoch (MJD 59523). All of these secondary spectra only show a fraction of their full Nyquist range.}
    \label{fig:all_spec1}
\end{figure}
\end{landscape}

\begin{landscape}
\centering
\begin{figure}
    \centering
    \includegraphics[width=1.25\textwidth]{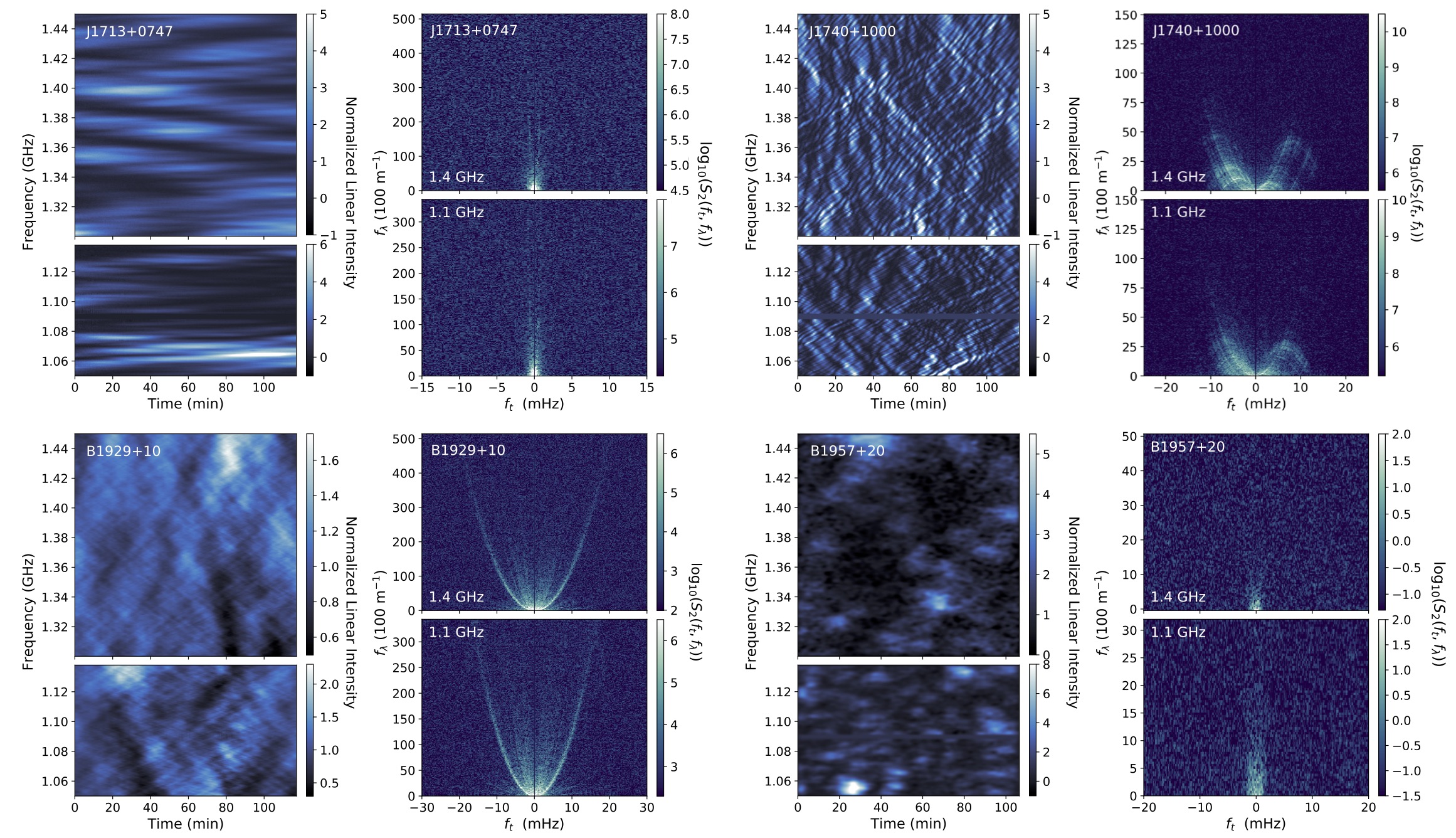}
    \caption{Dynamic and secondary spectra for (top left to bottom right): PSRs J1713$+$0747, J1740$+$1000, B1929$+$10, and B1957$+$20. Spectra were formed using the same methods as Figure~\ref{fig:all_spec1}, except a subintegration time of 6.4 s and frequency resolution of 1 MHz was used for B1957$+$20 due to its low S/N. The dynamic spectrum shown here for B1957$+$20 was further smoothed by a factor of 3 in frequency and time using a Gaussian kernel, for improved demonstration of the scintles (but the secondary spectrum was formed from the unsmoothed dynamic spectrum).}
    \label{fig:all_spec2}
\end{figure}
\end{landscape}

We define the arc curvature $\eta$ in frequency-independent coordinates $f_\lambda = \eta f_t^2$ \citep[e.g.][]{reardon2020}:
\begin{equation}\label{eq:eta}
    \eta = \frac{Ds(1-s)}{2V_\perp^2 {\rm cos}^2\psi},
\end{equation}

\noindent where $\psi$ {arises from the dot product between $\pmb{\theta}_{i}$ and $\pmb{V_\perp}$ (Eq.~\ref{eq:ft}) and} is the angle between {the screen's effective velocity} $\pmb{V}_\perp$ and the vector direction of points $\pmb{\theta}_{i}$ in the scattered image, if the image is anisotropic \citep{walker2004}. {If the image is isotropic then $\rm cos\ \psi = 1$}.  

{For comparison to previous studies,} the frequency-independent curvature $\eta$ may be converted to the curvature that would be measured in $f_t-f_\nu$ coordinates ($\eta_\nu$) simply using $\eta = c\eta_\nu/\lambda^2$. For $\eta_\nu$ in s$^3$ and $\nu$ in GHz, this gives $\eta \approx 33\times10^2\ {\rm m^{-1} mHz^{-2}} \times (\eta_\nu/{\rm s^3}) \times (\nu/{\rm GHz})^2$. 

The above discussion tacitly assumes that scintillation arises from the mutual interference of an angular spectrum of plane waves. An alternative description based on the Fresnel-Kirchoff diffraction integral nonetheless yields the same expressions given in Eq.~\ref{eq:ft}-\ref{eq:fnu} \citep{walker2004}. Scintillation arc theory based solely on refraction leads to these same expressions but refers to interference between multiple images of the pulsar, rather than points in a single scattered image \citep[e.g.][]{pen2014}. It is clear that the manifestation of parabolic arcs in the secondary spectrum is geometric and therefore generic; however, the power distribution along these arcs is not generic because it depends on the shape of the scattered image. In Section~\ref{sec:screen_constraints} we assess the distribution of power within observed secondary spectra, in the context of interstellar electron density fluctuations.

\subsection{Requirements to Detect Scintillation Arcs from Pulsar Bow Shocks}\label{sec:bow_shock_requirements}

If pulsar bow shocks produce scintillation arcs, then there is an optimal range of both pulsar and observing parameters to detect them. In this section we demonstrate that our observations meet these requirements in principle, barring S/N constraints.
A scintillation arc is only detectable if the highest point on the arc exceeds at least one sample in $f_\nu$. Assuming the arc fills the entire Nyquist range $[-f_{t,\rm Ny},+f_{t,\rm Ny}]$, then we require $f_\nu(f_{t,\rm Ny}) > \Delta f_\nu$, where $f_{t,\rm Ny} = 1/(2\Delta t)$ for a subintegration time $\Delta t$ and $\Delta f_\nu = 1/B$ for a total frequency bandwidth $B$. {The subintegration time $\Delta t$ corresponds to the time resolution of the dynamic spectrum, which is typically several seconds so that multiple pulses can be averaged together to achieve high S/N (see Section~\ref{sec:analysis}).} The parabola $f_\nu(f_{t,\rm Ny}) = \eta_\nu(f_{t,\rm Ny})^2$ yields a minimum detectable arc curvature 
\begin{equation}\label{eq:etamin1}
    \eta_{\nu,\rm min} = 4(\Delta t)^2/B,
\end{equation}
{or in the equivalent wavelength-derived curvature, $\eta_{\rm min} = 4c(\Delta t)^2/B\lambda^2$.} To further solve for the minimum detectable screen distance $s$, we must consider the relative velocities of the pulsar, screen (or bow shock), and observer. For a screen at the bow shock, $V_{\rm scr\perp} \lesssim V_{\rm psr\perp}$. Assuming an observer velocity much smaller than $V_{\rm psr\perp}$ and that $V_{\rm psr\perp} - V_{\rm scr\perp} \equiv \epsilon V_{\rm psr\perp}$, the screen's effective transverse velocity (Equation~\ref{eq:vperp}) reduces to $\pmb{V}_\perp \approx (\epsilon - s)\pmb{V}_{\rm psr\perp}$. For a bow shock, $s\ll0.1$ and $\epsilon\lesssim1$, and Equation~\ref{eq:eta} reduces to 
\begin{equation}\label{eq:etamin2}
    \eta_{\nu,\rm min} = Dsc/2\nu^2\epsilon^2V_{\rm psr\perp}^2,
\end{equation}
assuming cos$^2\psi=1$ for simplicity. Equations~\ref{eq:etamin1} and \ref{eq:etamin2} thus yield a minimum detectable screen distance
\begin{equation}
    s_{\rm min} = \frac{8\epsilon^2}{c}\frac{(\nu V_{\rm psr\perp} \Delta t)^2}{DB},
\end{equation}
which in physical units is
\begin{equation}\label{eq:dslmin}
    d_{\rm sl,min} = sD \approx {1.8\ \rm au} \times \frac{(\epsilon \nu_{\rm GHz} V_{\rm psr,100} \Delta t)^2}{B_{\rm GHz}}
\end{equation}
evaluated for frequencies in GHz, $\Delta t$ in seconds, and $V_{\rm psr\perp}$ in units of 100 km/s. If the scintillation arc only extends to a fraction $\kappa$ of the Nyquist range in $f_t$, then $d_{\rm sl,min}$ will be larger by a factor $\kappa^2$ in the denominator of Equation~\ref{eq:dslmin}. 

We thus find that fast subintegration times, large bandwidths, and lower velocity objects are most favorable for detection of arcs from pulsar bow shocks, assuming these arcs are high enough intensity to be detected. Low observing frequencies ($<1$~GHz) are likely less favorable, as arcs are generally observed to become increasingly diffuse at lower observing frequencies \citep{wu2022}.

\begin{figure}
    \centering
    \includegraphics[width=0.48\textwidth]{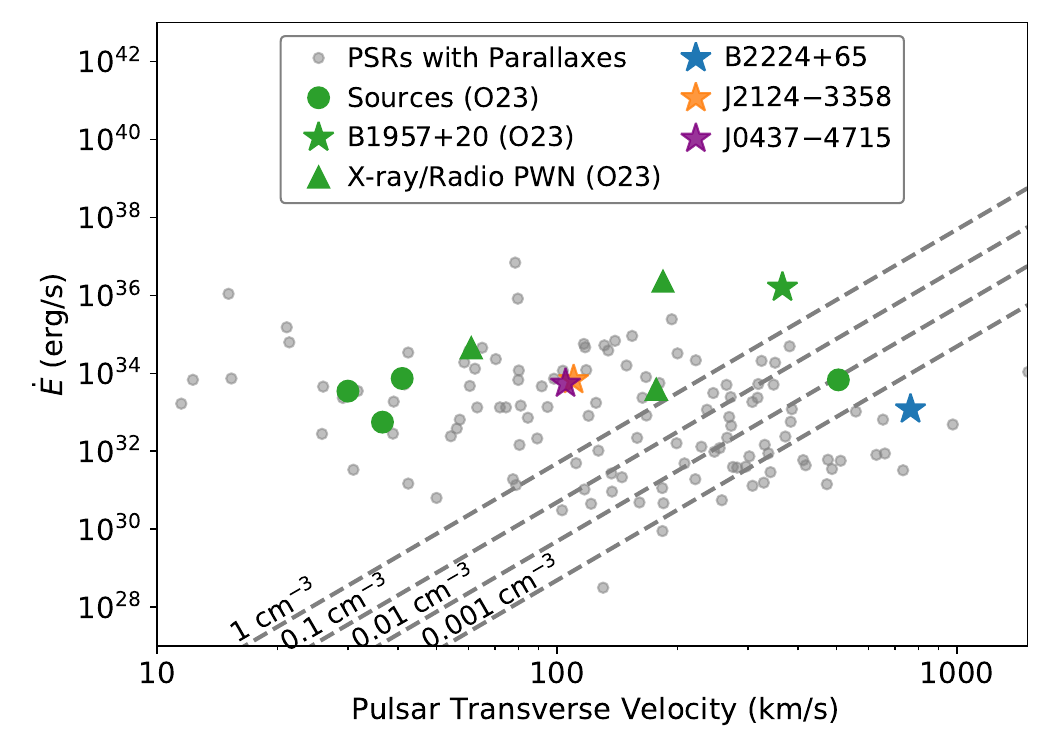}
    \caption{Spin-down luminosity $\dot{E}$ vs. transverse velocity $V_{\rm psr\perp}$, along with bow shock detectability for different values of the ISM electron density (dashed lines) assuming $\epsilon = 0.5$ (see Section~\ref{sec:bow_shock_requirements}). Pulsars with parallax measurements are shown in grey and our observed sources are shown in green and marked as O23 in the legend. Star symbols represent pulsars with confirmed H$\alpha$-emitting bow shocks. Sources with PWNe are shown as triangles. B2224$+$65 (Guitar Nebula), J2124$-$3358, and J0437$-$4715 are also highlighted as examples of other pulsars not observed in this work with known bow shocks. Pulsars that fall above the dashed lines meet the required $\dot{E}$ and $V_{\rm psr\perp}$ for scintillation arcs from their bow shocks to be resolved using the FAST observing parameters, assuming those arcs have sufficient S/N. Changing $\epsilon$ by {a factor of two} would have roughly the same impact on the minimum required $\dot{E}$ as changing the ISM density by one order of magnitude. }
    \label{fig:emin}
\end{figure}

Placing $d_{\rm sl,min}$ at the stand-off radius of the bow shock, we can solve for the range of spin-down luminosities and pulsar transverse velocities needed to resolve the bow shock as a scintillating screen. Assuming that the entire pulsar spin-down energy loss is carried away by the relativistic wind, the thin shell limit gives the bow shock stand-off radius:
\begin{multline}\label{eq:r0}
    R_0 = \sqrt{\frac{\dot{E}}{4\pi c \rho v_{*}^2}}\\
    \approx 225\ {\rm au} \times \bigg[\bigg(\frac{\dot{E}}{10^{33}\ {\rm erg\ s^{-1}}}\bigg) \bigg(\frac{n_H}{\rm cm^{-3}}\bigg)^{-1} \bigg(\frac{v_*}{100\ \rm km\ s^{-1}}\bigg)^{-2}\bigg]^{1/2},
\end{multline}
where $\dot{E}$ is the spin-down luminosity of the pulsar, $c$ is the speed of light, $v_*$ is the pulsar velocity, and $\rho = n_H \gamma_H m_H$ is the ISM density that depends on the number density of atomic hydrogen $n_H$, the cosmic abundance $\gamma_H$, and the mass of the hydrogen atom $m_H$ \citep{wilkin96,chatterjee2002}. Figure~\ref{fig:emin} shows the phase space of $\dot{E}$ vs. $V_{\rm psr \perp}$ for the pulsars observed, assuming $v_* = V_{\rm psr,\perp}$, and compared to different ISM electron densities $n_e$ calculated assuming $d_{\rm sl, min} = R_0$, $n_H \approx n_e$, and $\gamma_H = 1.37$. In principle, all of the pulsars observed in our study have high enough $\dot{E}$ and small enough $V_{\rm psr\perp}$ to yield a detectable scintillation arc from their bow shocks (if the bow shocks exist), for typical ISM densities. This statement does not account for the S/N of the arcs, which depends on the unknown scattering strength of the bow shocks. 

\section{Observations}\label{sec:obs}

\begin{table*}
    \renewcommand{\arraystretch}{1.15}
    \centering
    \begin{tabular}{ c c c c c c c c c c c}
    PSR & $l,b$ & $P$ & DM & $D$ & $V_{\rm psr\perp}$ & $\dot{E}$ & $S_{1400}$ & $R_0$ & Bow Shock/PWN & Epochs \\ 
     & (deg) & (ms) &  (pc cm$^{-3}$) &  (kpc) & (km/s) &  (erg s$^{-1}$) &  (mJy) & (au) & &  (MJD) \\  \hline \hline
    B0355$+$54 & $148, 0.8$ & 156 & 57.14 & $1.09_{-0.16}^{+0.23}$ & $61_{-9}^{+12}$ & $4.5\times10^{34}$ & 23 & 7900 & X-ray & 59509 \\ 
    B0919$+$06 & $225, 36$ & 430 & 27.29 & $1.21_{-0.16}^{+0.22}$ & $505\pm80$ & $6.8\times10^{33}$ & 10 & 370 & -- & 59499 \\
    B0950$+$08 & $228, 43$ & 253 & 2.97 & $0.262_{-0.005}^{+0.005}$ & $36.6\pm0.7$ & $5.6\times10^{32}$ & 100 & 1480 & Radio & 59500, 59523\\
    J1643$-$1224 & $5.7, 21$ & 4.62 & 62.41 & $0.763_{-0.099}^{+0.118}$ & $41^{+5}_{-4}$ & $7.4\times10^{33}$ & 4 & 4800 & -- & 59523, 59527\\
    J1713$+$0747 & $29, 25$ & 4.57 & 15.92 & $1.05_{-0.07}^{+0.06}$ & $30\pm2$ & $3.5\times10^{33}$ & 8 & 4500 & -- & 59509 \\
    J1740$+$1000 & $34, 20$ & 154 & 23.89 & 1.2 & 184 & $2.3\times10^{35}$ & 3 & 19000 & X-ray & 59510 \\
    B1929$+$10 & $47, -3.9$ & 226 & 3.18 & $0.361_{-0.009}^{+0.009}$ & $177_{-5}^{+4}$ & $3.9\times10^{33}$ & 29 & 805 & Radio, X-ray & 59515 \\
    B1957$+$20 & $59, -4.7$ & 1.61 & 29.12 & $2.57_{-0.48}^{+0.77}$ & $366^{+62}_{-98}$ & $1.6\times10^{35}$ & 0.3 & 8000 & H$\alpha$, X-ray & 59506 \\ \hline
    \end{tabular}
    \caption{The sample of observed pulsars and their properties. From left to right: Pulsar name, Galactic longitude and latitude, period, DM, distance, transverse velocity, spin-down luminosity, typical flux density at 1400 MHz, bow shock stand-off radius, wavelengths at which the bow shock or PWN has been detected, and observing epochs. Distances and transverse velocities are derived from VLBI parallax and proper motions (from \citealt{brisken2002,chatterjee2001, chatterjee2004,chatterjee2009,ding2023,romani2022}), except for J1740$+$1000 (see main text for details).
    For pulsars without confirmed H$\alpha$ bow shocks, the stand-off radii shown are rough estimates based on $\dot{E}$ and $V_{\rm psr\perp}$ assuming an ISM electron density of 0.1 cm$^{-3}$ and an inclination of $90^\circ$ (note for B1957$+$20, the resulting stand-off radius estimate would be $7880$ au, consistent with H$\alpha$ imaging; \citealt{romani2022}). All other parameters are retrieved from the ATNF catalogue \citep{atnf}.}
    \label{tab:source_list}
\end{table*}

We observed eight pulsars between October--December 2021 at FAST; the source list, pulsar properties, and observation dates are shown in Table~\ref{tab:source_list}. Our source list was primarily chosen based on the requirements described in Section~\ref{sec:bow_shock_requirements}, and a few of our sources are bright pulsars with previously detected scintillation arcs {(although specific arc properties were not factored into the pulsar selection)}. The sample includes both pulsars that have bow shocks previously observed in H$\alpha$ (B1957$+$20) or ram pressure confined PWNe observed in nonthermal radio or X-ray emission (B0355$+$54, B0950$+$08, and B1929$+$10), in addition to pulsars that do not have known bow shocks but do have spin-down luminosities and transverse velocities favorable for producing bow shocks and detectable scintillation arcs (see Section~\ref{sec:bow_shock_requirements}). None of these pulsars have supernova remnant associations, making a bow shock the most likely source of any scintillation arc with a screen distance very close to the pulsar.

Each source was observed for 2 hours at a single epoch, except for J1643$-$1224 and B0950$+$08, which were observed in two epochs separated by a few weeks. Data were recorded in filterbank format at a time resolution of 98 $\mu$s and a frequency resolution of $0.06$ MHz. FAST covers a frequency band of $1-1.5$ GHz, but bandpass roll-off at the upper and lower $10\%$ of the band yields an effective bandwidth from $1.05-1.45$ GHz. A noise diode injected an artificial modulated signal for one minute at the start and end of each observation, in order to verify gain stability and perform flux calibration.

\section{Data Reduction \& Analysis}\label{sec:analysis}

\subsection{Formation of Dynamic \& Secondary Spectra}
Dynamic spectra consist of the on-pulse intensity averaged over multiple pulses. After de-dispersion, the filterbank data were folded in 3.2 second long subintegrations using phase-connected timing solutions generated by \texttt{tempo} \citep{tempo}. {This subintegration time was chosen to provide sufficient coverage of very low-curvature arcs in the secondary spectrum, based on Eqs.~\ref{eq:etamin2}-\ref{eq:dslmin}; for $\Delta t = 3.2$ s and $B = 150$ MHz, $\eta_{\rm min} = 1.8\times10^{-3}$ m$^{-1}$ mHz$^{-2}$ at 1.4 GHz.} For B1957$+$20 a longer subintegration time of 6.4 seconds was used due to the pulsar's low S/N; {in this case, $\eta_{\rm min} = 7\times10^{-3}$ m$^{-1}$ mHz$^{-2}$}. The on-pulse signal was extracted from each folded subintegration using the phase range containing intensities within $90\%$ of the peak pulse intensity. The mean on-pulse flux density $S(t_i,\nu_i)$ for each subintegration $t_i$ and frequency channel $\nu_i$ was calibrated by subtracting the mean off-pulse flux density of each subintegration and dividing by the bandpass of the entire observation, {which was also calculated using off-pulse data}. In some epochs, the bandpass changed {slightly} over the 2-hour observing period {due to instrumental effects}, so multiple bandpasses were calculated for calibration. In all observations, wideband radio frequency interference (RFI) persisted between 1140 and 1300 MHz. We subsequently divided all dynamic spectra into two frequency bands: $1050-1140$ MHz and $1300-1450$ MHz. Transient RFI was masked and replaced with values interpolated from neighboring data points using a 2D Gaussian convolution kernel.

Before forming a secondary spectrum, each dynamic spectrum was interpolated onto a grid equispaced in wavelength $\lambda = c/\nu$, and a 2D Hanning window was applied to the outer edges of the dynamic spectrum to reduce sidelobe response in the secondary spectrum. The secondary spectrum was then formed from the squared magnitude of the 2D FFT of the dynamic spectrum: $S_2(f_t,f_\lambda) = |\tilde{S}(t,\lambda)|^2$. Gridding the dynamic spectrum in $\lambda$ yields a frequency-independent arc curvature across a single contiguous frequency band (see Equation~\ref{eq:eta}), and thus mitigates smearing of arc features in secondary spectra formed from broad bandwidths \citep{gwinn2019}. However the interpolation kernel used to resample the dynamic spectrum in $\lambda$ can have a dramatic effect on the secondary spectrum; e.g., we found that linear interpolation induces a non-linear drop-off in the (logarithmic) noise baseline of the secondary spectrum.
In order to ensure a flat noise baseline in subsequent analysis, we subtracted the mean logarithmic noise as a function of $f_\lambda$, calculated in a 5 mHz window at each edge of the secondary spectrum, and we note that inference of the power distribution along scintillation arcs can be biased by the choice of interpolation kernel if the shape of the off-noise baseline is not accounted for. The dynamic and secondary spectra for all eight pulsars are shown in Figures~\ref{fig:all_spec1} and \ref{fig:all_spec2}.

\subsection{Arc Identification and Curvature Measurements}
Scintillation arcs were identified by calculating the mean logarithmic intensity along parabolic cuts through the secondary spectrum for a range of arc curvatures. This procedure is also known as the generalized Hough transform \citep{ballard81,bhat2016}. For low arc curvatures (shallow arcs), the mean intensity was calculated out to a maximum $f_\lambda$  $\approx 10\%$ of the total Nyquist range of the secondary spectrum, in order to improve sensitivity to weaker arcs near the origin of the spectrum. For high arc curvatures, the mean intensity was calculated using a maximum $f_\lambda \approx 90\%$ of the total Nyquist range. In all but one case, curvatures correspond to a reference frequency of 1375 MHz because they were fit using the upper frequency band, which covered a larger contiguous bandwidth and contained less RFI. For B0919$+$06, curvatures were fit in the lower frequency band because it contained an additional arc that was not detected at 1375 MHz.

\begin{figure*}
    \centering
    \includegraphics[width=0.8\textwidth]{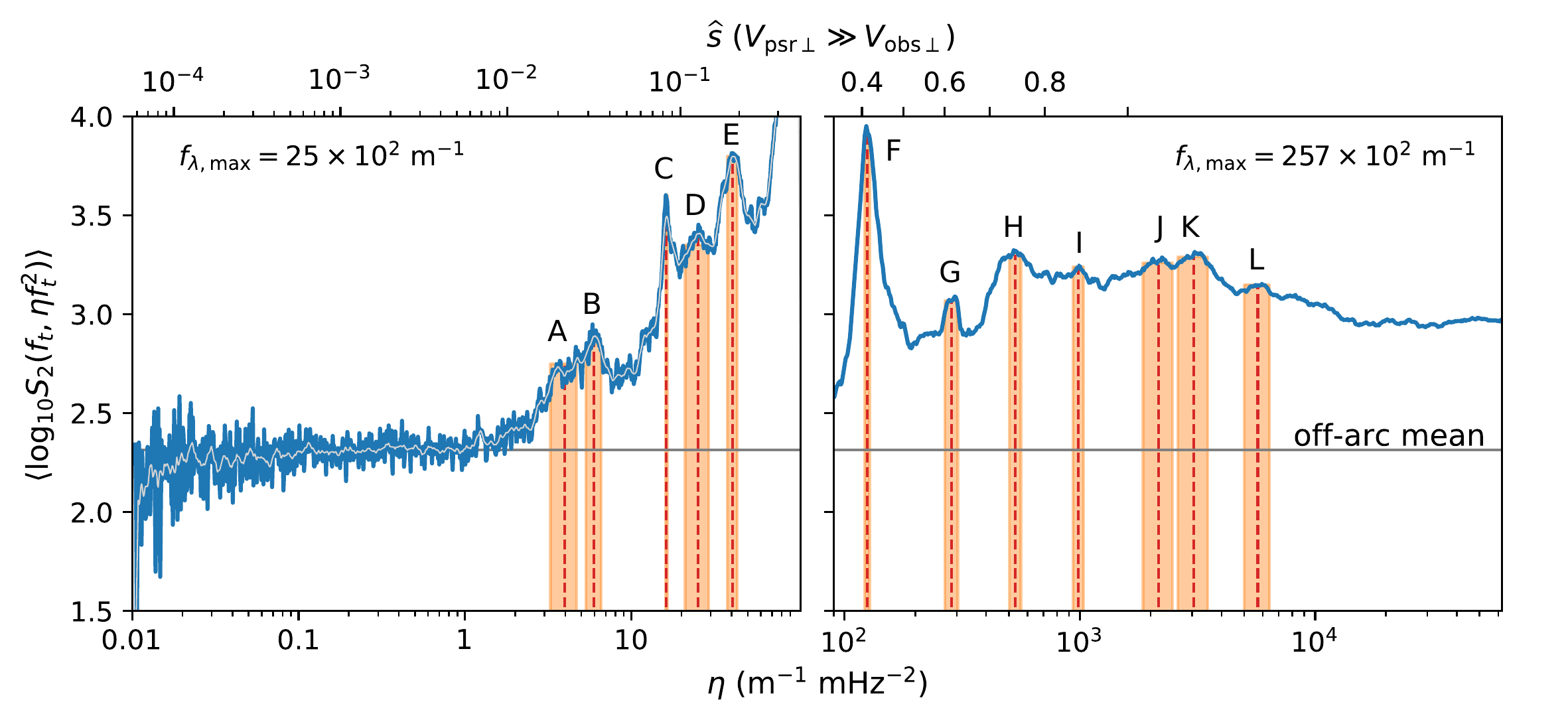}
    \caption{Mean logarithmic power along parabolic arcs in the secondary spectrum of B1929$+$10, as a function of arc curvature. The lefthand panel shows the mean power calculated up to a maximum delay of $2500\ \rm m^{-1}$, to improve the S/N of faint arcs at low curvatures. The righthand panel shows mean power calculated up to a maximum delay of $25700\ \rm m^{-1}$. In each panel the blue curve indicates the mean power along a parabolic cut through the secondary spectrum at a given curvature, while the red dashed lines and shaded orange regions respectively show the best-fit values and $1\sigma$ errors on the curvature of each arc in the secondary spectrum. The vertical axis scale is identical in the two frames and the intensity rise after feature E is part of feature F, but there is an apparent discontinuity between the frames because the averaging is done over different numbers of pixels. A light grey curve in the lefthand panel shows the mean power smoothed with a boxcar filter. The horizontal solid grey line in both panels shows the mean noise level in the secondary spectrum away from any scintillation arcs. The top axis indicates an approximate screen distance, assuming the pulsar's transverse velocity is much greater than those of the screen and observer.}
    \label{fig:B1929_parabfit}
\end{figure*}

An example of the resulting power distribution $\langle {\rm log}_{10} S_2 (f_t, \eta f_t^2) \rangle$ vs. curvature $\eta$ is shown in Figure~\ref{fig:B1929_parabfit} for B1929$+$10. Candidate arcs were identified as local maxima in the power distribution that were at least $1\sigma$ greater than their neighboring pixels, where $\sigma$ is the rms off-arc noise and each local maximum was required to span at least three pixels to avoid noise spikes. Each local maximum and its neighbors were then fit with an inverted parabola, and the value of $\eta$ at the peak of the fitted parabola was taken to be the best-fit curvature for the candidate arc. The associated error on $\eta$ was determined from the range within which the fitted, inverted parabola was $<1\sigma$ below its peak (where $\sigma$ again is the rms off-arc noise). Similar procedures have been used by, e.g., \cite{reardon2020} and \cite{mckee2022}. Candidate arcs were then sorted by their uniqueness; i.e., for candidates with the same $\eta$ to within the errors, only the highest S/N candidate was selected for the final set of arcs reported for each pulsar. 

In some cases (e.g., B0919$+$06, B0950$+$08, and B1957$+$20), the power distribution increases to a power maximum that extends over a broad range of $\eta$ values, corresponding to a ``bounded'' arc that contains diffuse power filling in the entire arc's extent in $f_t$. In these cases, the arc curvature is reported as a lower limit based on the curvature at which the mean power distribution reaches $95\%$ of its maximum. 

The methods described above assume that the noise in the power distribution follows Gaussian statistics; however, this is not generally true for very low curvatures because interpolating the dynamic spectrum onto an equi-spaced wavelength grid introduces correlated noise that has a correlation length greater than a few samples at low $f_\lambda$. While our procedure for identifying arcs does not directly account for this correlated noise, we note that all of our methods were repeated on secondary spectra formed from the original, radio frequency-domain dynamic spectra. We found no difference in the results other than a reduced precision in the arc curvature measurements, due to the frequency-dependent smearing of arc features.

\begin{figure*}
    \centering
    \includegraphics[width=0.9\textwidth]{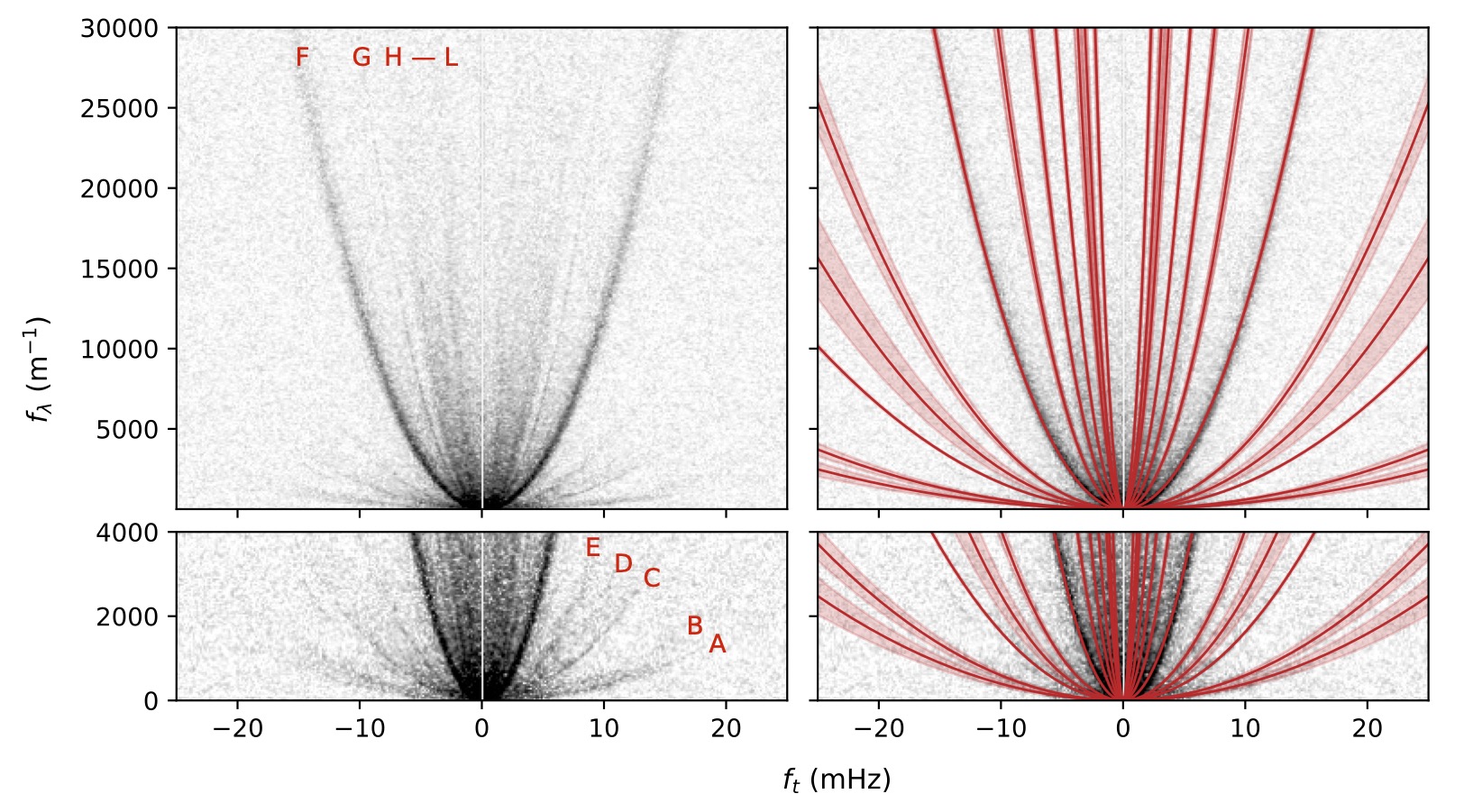}
    \caption{A demonstration of the best-fit curvatures for B1929$+$10. The lefthand panels show the secondary spectrum at 1.4 GHz for two ranges in $f_\lambda$, up to $3\times10^4\ \rm m^{-1}$ (top) and up to $4\times10^3\ \rm m^{-1}$ (bottom). The red curves and shaded regions in the righthand panels show the best-fit curvatures and $1\sigma$ errors overlaid on the secondary spectrum. Letters identify features of interest from the Hough transform (Figure~\ref{fig:B1929_parabfit}).}
    \label{fig:B1929_curvfit_demo}
\end{figure*}

Demonstrations of the best-fit curvatures compared to original secondary spectra are shown in Figures~\ref{fig:B1929_curvfit_demo}-\ref{fig:B0950_curvfit_demo} for B1929$+$10, B0355$+$54, and B0950$+$08, which display the range of arc traits observed.

\section{Results}\label{sec:results}

Scintillation arcs were detected for all of the pulsars in this study. The secondary spectra shown in Figures~\ref{fig:all_spec1}-\ref{fig:all_spec2} reveal diverse scintillation characteristics, ranging from thin, highly defined arcs (e.g. B1929$+$10, J1713$+$0747) to diffuse, broad arcs (B0355$+$54, J1643$-$1224), filled-in arcs (B0919$+$06, B0950$+$08, B1957$+$20), and in one case dramatic reverse arclets (J1740$+$1000). We find a number of additional arcs beyond those previously reported for pulsars in the dataset, including low-curvature, truncated arcs for B1929$+$10, B0355$+$54, B0919$+$06, and B0950$+$08. B1929$+$10 notably shows an extremely large concentration of arcs, discussed further below. 

For each pulsar we report arc curvatures, and infer estimates of the fractional screen distance $s$ using Equations~\ref{eq:vperp} and \ref{eq:eta}. The effective screen velocity $V_\perp$ was calculated using the pulsar's transverse velocity (Table~\ref{tab:source_list}) and Earth's transverse velocity relative to the LOS based on the \cite{moisson2001} ephemeris model implemented in \texttt{astropy} and \texttt{scintools} \citep{astropy2022,reardon2020}. The estimated error in the Earth velocity term is negligible compared to the uncertainty in the pulsar velocity term. In general, the screen velocity $V_{\rm scr\perp}$ and the angle of anisotropy $\psi$ are not independently measurable, and scintillation studies use multi-epoch measurements of arc curvature variations to break the degeneracy of these parameters when inferring $s$ \citep[e.g.][]{main2020,reardon2020}. 
Lacking enough multi-epoch measurements to do such an analysis, we instead make fiducial estimates of $s$ by assuming $ V_{\rm scr \perp} \ll V_{\rm psr\perp}, V_{\rm obs \perp}$ (where appropriate) and $\psi = 0^\circ$. Figure~\ref{fig:eta_vs_s} shows $\eta$ vs. $s$ for different values of $\psi$ and $V_{\rm scr\perp}$, using B1929$+$10 as an example ($V_{\rm psr\perp} = 177$ km/s, $V_{\rm obs\perp} = 16$ km/s). Larger values of $V_{\rm scr\perp}$ and $\psi$ both result in larger $\eta$ for a given $s$, although $V_{\rm scr\perp}$ has the largest impact for screens near the observer ($s\gtrsim0.8$) when it is comparable to or larger than $V_{\rm obs\perp}$. In most cases, assuming $\psi = 0^\circ$ and $ V_{\rm scr \perp} \ll (V_{\rm psr\perp}, V_{\rm obs \perp})$ thus yields an upper limit on $s$. However, we note several instances below where the measured arc curvature requires either larger $\psi$ and/or larger $V_{\rm scr\perp}$, and in Section~\ref{sec:bow_shock_screens} we consider potential bow shock screens with larger $V_{\rm scr\perp}$. In some cases, the depth of the intensity valley within an arc may also be an indicator of anisotropy (see e.g. Appendix B of \citealt{reardon2020}). Figure~\ref{fig:eta_vs_s} demonstrates that Equation~\ref{eq:eta} technically yields two possible values of $s$ for a given $\eta$. For several pulsars in our dataset, $V_{\rm psr\perp} \gg V_{\rm obs\perp}$ and we can ignore the solution at $s\approx1$. The results for each pulsar are elaborated below and briefly compared to previous relevant observations. 

\subsection{B0355$+$54}

\begin{figure*}
    \centering
    \includegraphics[width=0.9\textwidth]{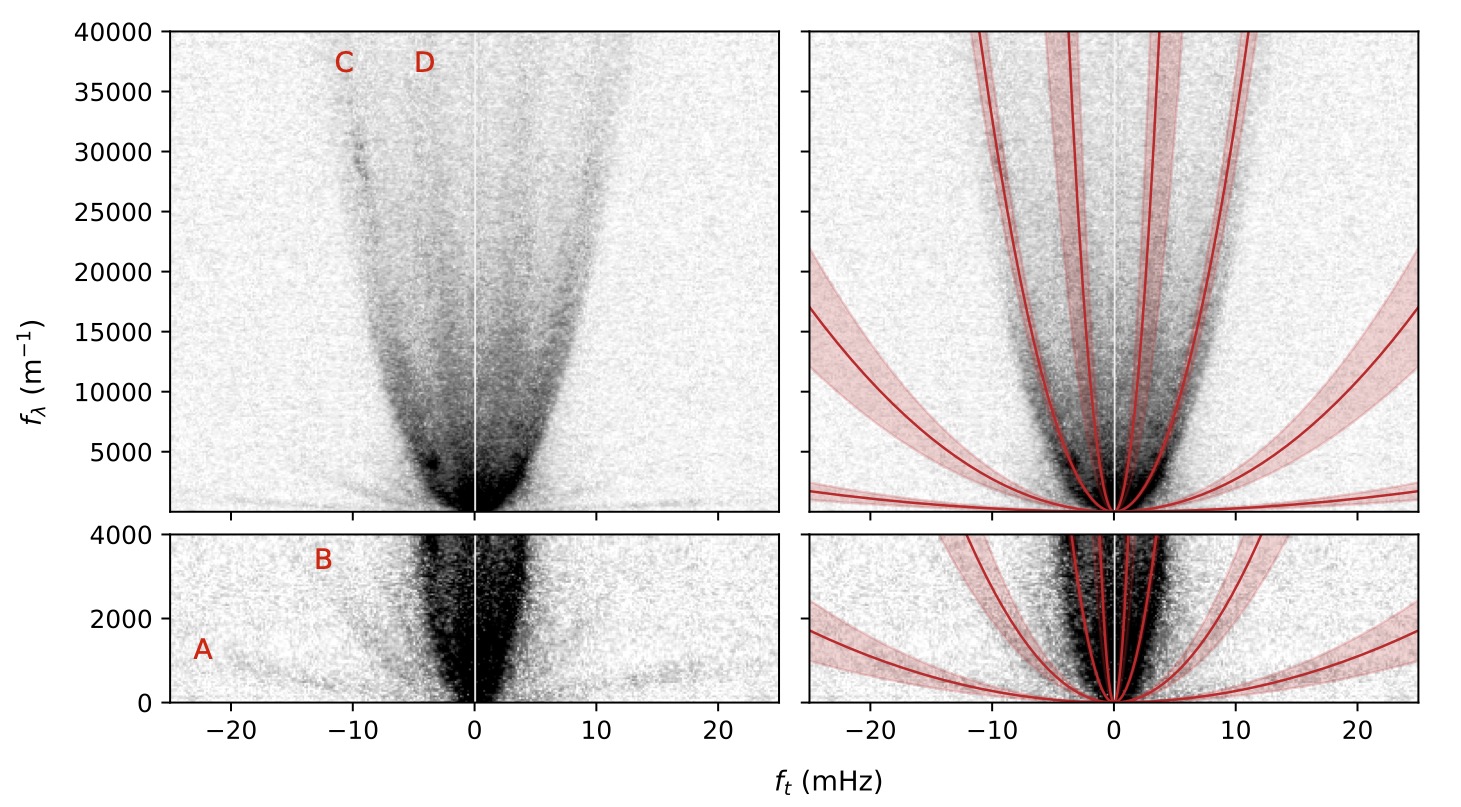}
    \caption{A demonstration of the best-fit curvatures for B0355$+$54 at 1.4 GHz (similar to Figure~\ref{fig:B1929_curvfit_demo}). In this case, four arcs are detected in the power summation procedure (see Section~\ref{sec:analysis}), although a third low curvature arc may be marginally visible in the secondary spectrum at negative $f_t$.}
    \label{fig:B0355_curvfit_demo}
\end{figure*}

B0355$+$54 displays four scintillation arcs whose curvatures are shown in Table~\ref{tab:B0355results}.
The best-fit curvatures are also shown overlaid on the original secondary spectrum in Figure~\ref{fig:B0355_curvfit_demo}. A fifth, one-sided arc at a curvature of about 60 m$^{-1}$ mHz$^{-2}$ is marginally visible in Figure~\ref{fig:B0355_curvfit_demo} but was not considered significant based on our detection criteria (Section~\ref{sec:analysis}). The arcs have a smooth and diffuse visual appearance, although Arc C (curvature $327\pm38$ m$^{-1}$ mHz$^{-2}$) contains multiple power enhancements spanning $\lesssim 25\times10^3$ m$^{-1}$ in $f_\lambda$. 

For B0355$+$54, $V_{\rm obs\perp} \approx 25$ km/s, a significant fraction of the pulsar's transverse velocity $V_{\rm psr\perp} = 61^{+12}_{-9}$ km/s \citep{chatterjee2004}. We therefore estimate a range of screen locations for each arc, considering the possibility that some screens may be either closer to the pulsar or closer to the observer, and that the screen orientation $\psi$ is not constrained. The estimated ranges of $s$ are shown in Table~\ref{tab:B0355results}. For the lowest curvature arc, the near-pulsar solution is $s \leq (6\pm3)\times10^{-4}$ for $\psi \geq 0^\circ$, where error bars correspond to the propagated uncertainties on the arc curvature. For $D = 1.09$ kpc \citep{chatterjee2004} this $s$ corresponds to a physical distance $\leq 0.98$ pc $\approx 2\times10^5$ au from the pulsar, and the screen could be substantially closer if $\psi > 0^\circ$. Conversely, the near-observer solution for this arc yields $s > 0.9998$, which would correspond to a screen extremely close ($<1$ pc) from the observer (after accounting for uncertainties in the pulsar distance). Previous studies of this pulsar have observed single scintillation arcs with variable power enhancements over hour to month timescales \citep{xu2018_b0355,wang2018,stinebring2007}. \cite{wang2018} measure an arc curvature of $0.0216$~s$^{3}$ at 2.25 GHz, equivalent to $361$~m$^{-1}$~mHz$^{-2}$, which is broadly consistent with the curvature of Arc C. 

\begin{table}
    \centering
    \begin{tabular}{c | c | c | c}
     Feature & Curvature & Fractional Screen & Assumed \\ 
     Identifier  &  (m$^{-1}$ mHz$^{-2}$) & Distance & $\psi$ (deg) \\ \hline \hline
     \multirow{2}{*}{A} & \multirow{2}{*}{$2.7\pm1.2$} & $s\leq 9\times10^{-4}$ &$\psi \geq 0$\\ 
     & & $s>0.9998$ & $\psi \geq 0$ \\ \hline 
     \multirow{2}{*}{B} & \multirow{2}{*}{$27\pm8$} & $(2\lesssim s \lesssim 8)\times10^{-3}$ & $0 \leq \psi \leq 45$ \\
     & & $s>0.9986$ & $\psi \geq 0$ \\ \hline
     \multirow{2}{*}{C} & \multirow{2}{*}{$327\pm38$} & $0.031 \lesssim s \lesssim 0.08$ & $0 \leq \psi \leq 45$ \\
     & & $s> 0.985$ & $\psi \geq 0$ \\ \hline 
     \multirow{2}{*}{D} & \multirow{2}{*}{$2869\pm1603$} & $0.1 \lesssim s \lesssim 0.5$ & $0 \leq \psi \leq 45$ \\
     & & $0.91 \lesssim s \lesssim 0.97$ & $0 \leq \psi \leq 45$\\
    \end{tabular}
    \caption{Scintillation arc curvatures and fractional screen distances $s$, inferred at 1.4 GHz for B0355$+$54. A range of screen distances is shown for both near-pulsar and near-Earth solutions based on fiducial assumptions of $\psi$ shown in the table, and assuming $V_{\rm scr\perp} \ll V_{\rm psr\perp}$. The uncertainties in the arc curvature were propagated into uncertainties on $s$, to determine $\pm1\sigma$ upper and lower limits that give the ranges of $s$ shown. Note that increasing $V_{\rm scr\perp}$ would decrease $s$. Assuming a value or range of $\psi$ does not break the twofold degeneracy relating $\eta$ and $s$, due to the significant observer velocity contribution along this LOS; in future, multi-epoch curvature measurements will be needed to uniquely determine $s$.}
    \label{tab:B0355results}
\end{table}

\subsection{B0919$+$06}

Two arcs are detected for B0919$+$06, one shallow and highly truncated arc at $\eta = 13\pm2$ m$^{-1}$ mHz$^{-2}$ and a diffuse, filled in arc with $\eta \geq 69$ m$^{-1}$ mHz$^{-2}$ (1.1 GHz). Two marginal arcs may also be present at $0.1$ and $2$ m$^{-1}$ mHz$^{-2}$, but were just $1\sigma$ above the noise baseline and did not meet the detection threshold criteria. Due to this pulsar's extremely large transverse velocity, $V_{\rm psr \perp} = 505\pm80$ km/s \citep{chatterjee2001}, we ignore the near-observer solutions for screen distance $s$, and subsequently find $0.07 \lesssim s \lesssim 0.2$ ($0^\circ \leq \psi \leq 45^\circ$) and $0.3 \lesssim s \lesssim 0.5$ ($0^\circ \leq \psi \leq 45^\circ$). For $D = 1.21$ kpc \citep{chatterjee2001}, these screens span physical distances $\approx0.6$ to $1.1$ kpc from the observer. 
 
Scintillation arcs have been previously observed for this pulsar at different radio frequencies by \cite{stinebring2001}, \cite{putney2006}, and \cite{stinebring2007}. \cite{chatterjee2001} argued that the scintillation velocity is consistent with a scattering screen $\sim 250$ pc from observer. The two arcs reported by \cite{putney2006} are broadly consistent with the arcs reported here.

\subsection{B0950$+$08}

\begin{figure*}
    \centering
    \includegraphics[width=0.75\textwidth]{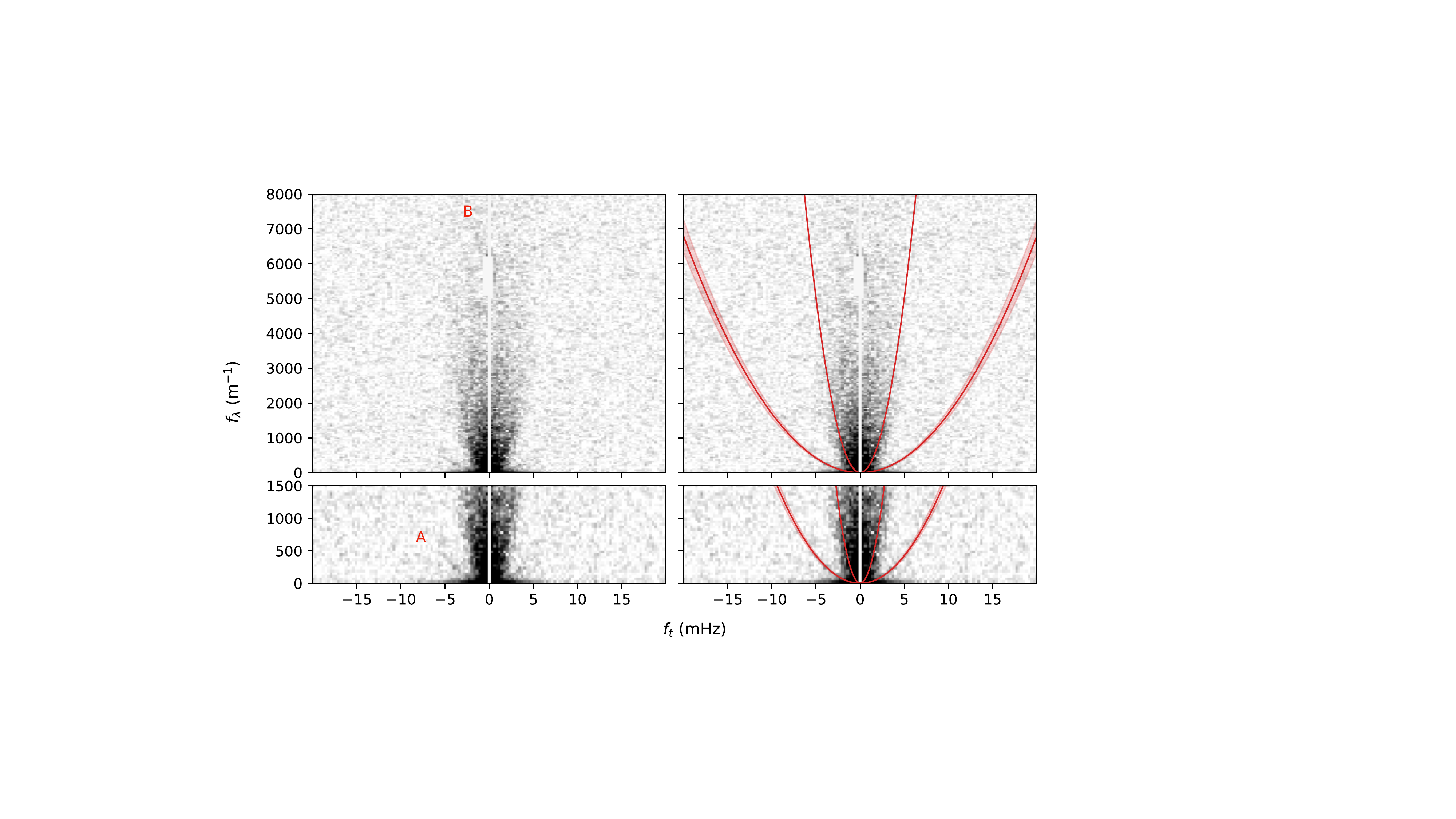}
    \caption{A demonstration of the best-fit curvatures for B0950$+$08 at 1.4 GHz (similar to Figure~\ref{fig:B1929_curvfit_demo}), for the observation on MJD 59523. In this case, one thin and highly truncated arc is detected at low curvature, and a broader distribution of power within a parabolic boundary is detected at higher curvature. In the latter case, a lower limit is reported on the arc curvature, indicated here by the solid red line.}
    \label{fig:B0950_curvfit_demo}
\end{figure*}

\begin{figure}
    \centering
    \includegraphics[width=0.45\textwidth]{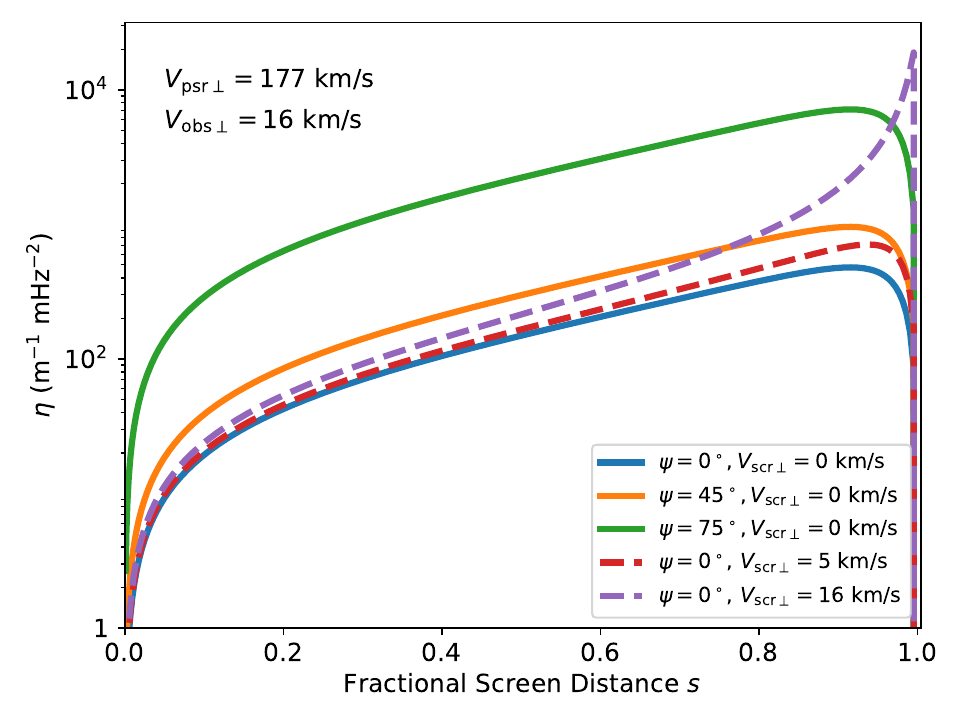}
    \caption{Arc curvature $\eta$ vs. screen distance $s$ for different values of the screen velocity $V_{\rm scr\perp}$ and angle of anisotropy $\psi$, assuming $V_{\rm psr\perp} = 177$ km/s and $V_{\rm obs\perp} = 16$ km/s (as for B1929$+$10). Larger $\psi$ and $V_{\rm scr\perp}$ both increase the possible values of $\eta$ for a given $s$, and $V_{\rm scr\perp}$ has the largest effect on $\eta$ near $s\approx 1$ when it is comparable to or greater than the observer velocity term.}
    \label{fig:eta_vs_s}
\end{figure}

B0950$+$08 displays remarkably similar scintillation arcs to B0919$+$06: one thin, highly truncated arc is detected at $\eta = 17\pm1$ m$^{-1}$ mHz$^{-2}$, in addition to a broad filled in arc at $\eta \geq 200$ m$^{-1}$ mHz$^{-2}$. The best-fit curvatures are shown overlaid on the secondary spectrum in Figure~\ref{fig:B0950_curvfit_demo}. Due to its low transverse velocity ($V_{\rm psr \perp} = 36.6\pm0.7$ km/s; \citealt{brisken2002}), we estimate two ranges of $s$ for each arc:
$0.003 \lesssim s \lesssim 0.006$ and $0.03 \lesssim s \lesssim 0.06$ (assuming $0^\circ \leq \psi \leq 45^\circ$) for the near-pulsar solutions of the two arcs, respectively, and $s \gtrsim 0.997$ for the near-observer solutions. The shallowest arc corresponds to a screen either $<1.6$ pc from the pulsar or $<0.3$ pc from the observer, regardless of $\psi$, and could be twice as close to either the pulsar or observer if $\psi \gtrsim 45^\circ$. The second screen is either $0.8$ pc or $\sim 240 - 260$ pc from the observer, depending on $\psi$ and the uncertainty in the pulsar distance. 

\cite{wu2022} observed a single scintillation arc for B0950$+$08 with LOFAR in 2016, with curvature $\eta_\nu = 4.8\pm0.7$ s$^3$ at 150 MHz, equivalent to $\eta \approx 356$ m$^{-1}$ mHz$^{-2}$ at 1.4 GHz, and estimated a screen distance of $230\pm35$ pc for $\psi = 0^\circ$ by ignoring the velocity of the screen. Our results for Arc B ($\eta \geq 200$ m$^{-1}$ mHz$^{-2}$) are broadly consistent, suggesting that the scattering screen responsible for this arc may have persisted for over five years or longer. \cite{smirnova2014} used VLBI to resolve the scattered image of B0950$+$08 at 324 MHz and found evidence for scattering in two layers at distances $\sim 10$ pc and $26 - 170$ pc from the observer, neither of which appear to be consistent with our observations.

\subsection{J1643$-$1224}\label{sec:J1643}

J1643$-$1224 shows a single, very broad scintillation arc with $\eta = 5089 \pm 3536$ m$^{-1}$ mHz$^{-2}$ (1.4 GHz). In this case, $V_{\rm obs\perp} \approx 0.5 V_{\rm psr\perp}$ \citep{ding2023} and the single-epoch measurement combined with large curvature uncertainties yield an extremely large range of possible screen distances; e.g., for $\psi = 45^\circ$ $s \approx 0.6^{+0.3}_{-0.5}$. \cite{mall2022} previously measured the scintillation arc curvature as it varied over a five-year period and found a best-fit screen distance $\approx 114 - 223$ pc from the observer, consistent with the distance to a foreground HII region \citep{harveysmith2011,ocker2020}. Our scintillation arc measurement is broadly consistent with the \cite{mall2022} result. While our secondary spectrum has a Nyquist limit of about $(0.12\ \rm MHz)^{-1} \approx 8$ $\mu$s, the arc observed by \cite{mall2022} extends up to $f_\nu \approx 20-30$ $\mu$s, implying that our observation is sensitive to only a small portion of the full scintillation arc.

\subsection{J1713$+$0747}

Visually, the secondary spectrum for J1713$+$0747 shows scintillation arc structure on two scales: one high-contrast, thin arc that rises above an exterior, diffuse region of power closer to the origin. The Hough transform detects three arcs at curvatures of $1184\pm33$, $8269\pm2491$, and $38531\pm13640$ m$^{-1}$ mHz$^{-2}$ (1.4 GHz). Similar to J1643$-$1224, this pulsar's transverse velocity is just twice $V_{\rm obs\perp}$ \citep{chatterjee2009}, and the near-pulsar and near-observer screens are indistinguishable with the data in hand. For the shallowest arc, we find either $s \lesssim 0.07$
or $s \geq 0.96$ for $\psi \geq 0^\circ$, corresponding to physical distances $\lesssim 70$ pc from the pulsar or $\lesssim 40$ pc from the observer. However, for the pulsar and observer velocity configuration of this LOS, the maximum arc curvature yielded by Equation~\ref{eq:eta} for $\psi = 0^\circ$ is just $5820$ m$^{-1}$ mHz$^{-2}$, which is too small to explain the curvatures of the two steepest arcs in the secondary spectrum. We thus find that larger $\psi$ and/or larger $V_{\rm scr\perp}$ are required for the higher curvature arcs along this LOS. Assuming $\psi = 45^\circ$ and $V_{\rm scr\perp} = 0$ km/s, we find screen solutions at either $s = 0.3\pm0.1$ (near-pulsar) or $s = 0.8\pm0.1$ (near-observer) for the arc at curvature $8269\pm2491$ m$^{-1}$ mHz$^{-2}$. For the highest curvature ($38531\pm13640$ m$^{-1}$ mHz$^{-2}$), assuming $V_{\rm scr\perp} = 0$ km/s requires $\psi > 60^\circ$, and we find $s \approx 0.6^{+0.2}_{-0.3}$ for $\psi = 65^\circ$. A scintillation arc has been measured for this pulsar only once before \citep{main2023b}, making this LOS a prime target for dedicated follow-up observations. 

\subsection{J1740$+$1000}\label{sec:J1740}

J1740$+$1000 is the only pulsar in the dataset to display well-defined reverse arclets, which could be due to interference between discrete sub-components and/or a high degree of anisotropy in the scattered image. \cite{sprenger2021} and \cite{baker2022} developed a method to measure the curvature of such reverse arclets, assuming a 1D scattered image, by linearizing the secondary spectrum so that the forward arc and reverse arclets all lie along straight lines through a transformation of $f_t - f_\nu$ space. Application of this ``$\theta-\theta$ transform'' to the J1740$+$1000 secondary spectrum yields a best-fit curvature of $\eta = 72\pm5$ m$^{-1}$ mHz$^{-2}$ (see Appendix~\ref{app:theta-theta}). Although J1740$+$1000 lacks parallax and proper motion measurements, its scintillation speed implies a transverse velocity of $\approx 184$ km/s, which could be much larger based on its location far above the Galactic plane \citep{mclaughlin2002}. NE2001 \citep{ne2001} and YMW16 \citep{ymw16} both predict a distance of 1.2 kpc for this pulsar. We estimate a screen distance $s \leq 0.13$ or 160 pc from the pulsar (for $\psi \geq 0^\circ$), but a parallax distance is required to obtain a more accurate estimate of the screen distance in physical units. Scintillation arcs have not been previously reported for this pulsar, although \cite{rozko2020} observed a turnover in the pulsar spectrum below 300 MHz that may be due to interstellar absorption along the LOS. 

\subsection{B1929$+$10}

This pulsar shows the largest concentration of arcs among our observations. The Hough transform detection criteria yield 12 arc candidates; however, the three highest-curvature candidates (J--L in Figure~\ref{fig:B1929_parabfit} and Table~\ref{tab:B1929results}) are all superposed on a broad power distribution and we remain agnostic as to whether these are three independent arcs tracing distinct scattering screens. The best-fit curvatures are shown in Table~\ref{tab:B1929results}.

Table~\ref{tab:B1929results} also shows estimates of the screen distance for each arc, assuming $V_{\rm psr\perp} \gg (V_{\rm obs\perp}, V_{\rm scr\perp})$, a reasonable assumption given $V_{\rm psr\perp} = 177$ km/s \citep{chatterjee2004}. Allowing different values of $\psi$ for each screen can yield overlapping screen distances for different arcs. In Table~\ref{tab:B1929results} we show two possible screen distances for $\psi = 0^\circ$ and $\psi=45^\circ$. Presuming that distinct arcs are observed when the scattering screens do not overlap, then some combination of $\psi$ (and $V_{\rm scr\perp}$) is required that yields unique values of $s$ for each arc. In practice, disentangling the degeneracy between $\psi$ and $V_{\rm scr\perp}$ for each of $>9$ arcs will require many repeated observations. Nonetheless, our fiducial estimates suggest that the LOS to B1929$+$10 contains a high filling fraction of scattering material, with screens spanning $\sim90\%$ of the 361 pc path-length to the pulsar. In addition, the large curvatures of arcs H, I, and candidate arcs J--L all appear to require $\psi \gtrsim 45^\circ$ (and/or velocity vector alignment such that $|\mathbf{V_{\rm eff}}|$ is small, which could result from larger $V_{\rm scr\perp}$). 

\begin{figure}
    \centering
    \includegraphics[width=0.46\textwidth]{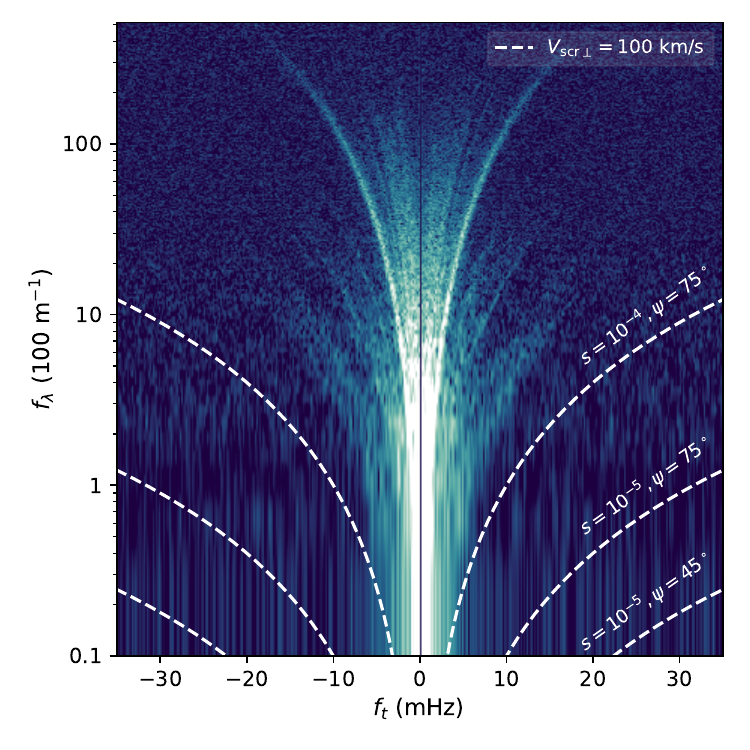}
    \caption{The secondary spectrum for B1929$+$10 at 1.4 GHz, viewed on a logarithmic scale in $f_\lambda$. The white dashed curves show three examples of where arcs would fall in the secondary spectrum for different combinations of $s$, $\psi$, and $V_{\rm scr\perp}$. Nominal estimates of the bow shock stand-off radius for B1929$+$10 would suggest $s\sim10^{-5}$, although $s$ could be larger depending on the inclination angle of the bow shock relative to the pulsar LOS. Regardless, the scattering screen at the shock would need to have large $\psi$ and $V_{\rm scr\perp}$ to explain the lowest curvature arc detected here.}
    \label{fig:B1929_zoom}
\end{figure}

Up to three distinct scintillation arcs have been observed for B1929$+$10 in the past \citep{putney2006,cordes2006,fadeev2018,yao2020,wu2022}. The high sensitivity of FAST reveals numerous additional arcs, including low-curvature arcs (features A--E in Figures~\ref{fig:B1929_parabfit}-\ref{fig:B1929_curvfit_demo}) that precede the highest intensity Arc F. These low-curvature arcs were identified using the dynamic spectrum in the 1.4 GHz band. The secondary spectrum formed from the 1.1 GHz band only showed one arc where features A-B are, which is likely due to the smaller bandwidth of the 1.1 GHz dynamic spectrum. Figure~\ref{fig:B1929_zoom} shows the low-curvature arcs in enhanced detail. They are $\sim 100\times$ weaker than Arc F and confined to $f_\lambda < 30$ m$^{-1}$. The shallowest of these, Arc A, has a screen distance $s < 0.027$, equivalent to $<9.7$ pc from the pulsar. Increasing $\psi$ to $>45^\circ$ could bring the screen distance to within 1 pc of the pulsar. 

\cite{wu2022} find a single arc with a curvature of $3.0\pm0.1$~s$^3$ at 150 MHz, equivalent to $223\pm74$~m$^{-1}$~mHz$^{-2}$ and consistent with the curvature of Arc G. This arc curvature is also broadly consistent with the arc observed by \cite{fadeev2018} and \cite{yao2020}. \textit{A priori}, Arc F would be a plausible arc to associate with previous observations given that it contains the most power of any arc in our secondary spectrum. Follow-up observations that track the curvatures of all of the arcs reported here will confirm which have indeed been observed in prior studies.

{B1929$+$10 is the only pulsar in our sample that has shown conclusive evidence of tiny-scale atomic structure (TSAS) detected in HI absorption of the pulsar spectrum. \cite{stanimirovic2010} measured up to four distinct TSAS features in the pulsar spectrum, with spatial scales $\approx 6 - 45$ au based on the temporal variability of the HI absorption features. While the distances of the TSAS features could not be directly determined from the pulsar spectrum, \cite{stanimirovic2010} suggested that they are $\approx 106 - 170$ pc from the observer based on the similarity between the TSAS velocities and the velocity of NaI absorption features observed towards stars within $3^\circ$ of the pulsar LOS. These TSAS features could be related to the same physical processes that are responsible for the large concentration of scintillation arcs along this LOS.}

\begin{table}
    \centering
    \begin{tabular}{c | c | c | c}
     Feature & Curvature & Fractional Screen & Assumed \\
     Identifier & (m$^{-1}$ mHz$^{-2}$) & Distance & $\psi$ (deg) \\\hline \hline
     \multirow{2}{*}{A} & \multirow{2}{*}{$4.0\pm0.7$} & $0.022\pm0.005$ & 0 \\ 
     & & $0.011\pm0.002$ & 45 \\ \hline
     \multirow{2}{*}{B} & \multirow{2}{*}{$5.9\pm0.6$} & $0.032\pm0.003$ & 0 \\
     & & $0.016\pm0.002$ & 45 \\ \hline
     \multirow{2}{*}{C} & \multirow{2}{*}{$16.2\pm0.2$} & $0.085\pm0.001$ & 0 \\
     & & $0.0439\pm0.0004$ & 45 \\ \hline
     \multirow{2}{*}{D} & \multirow{2}{*}{$25\pm4$} & $0.13\pm0.02$ & 0\\
     & & $0.07\pm0.01$ & 45 \\ \hline
     \multirow{2}{*}{E} & \multirow{2}{*}{$41\pm3$} & $0.19\pm0.01$ & 0\\
     & & $0.105\pm0.006$ & 45 \\ \hline
     \multirow{2}{*}{F} & \multirow{2}{*}{$125\pm3$} & $0.448\pm0.007$ & 0\\
     & & $0.274\pm0.005$ & 45 \\ \hline
     \multirow{2}{*}{G} & \multirow{2}{*}{$285\pm13$} & $0.71\pm0.02$ & 0\\
     & & $0.49\pm0.02$ & 45 \\ \hline
     \multirow{2}{*}{H} & \multirow{2}{*}{$532\pm30$} & \multirow{2}{*}{$0.68\pm0.02$} & \multirow{2}{*}{45} \\
     & & & \\ \hline
     \multirow{2}{*}{I} & \multirow{2}{*}{$987\pm51$} & \multirow{2}{*}{$0.92\pm0.04$} & \multirow{2}{*}{45} \\
     & & &  \\ \hline
     \multirow{2}{*}{J} & \multirow{2}{*}{$2164\pm308$} & \multirow{6}{*}{$s\gtrsim0.9$} & \multirow{4}{*}{$\gtrsim 60$} \\
     & & & \\ 
     \multirow{2}{*}{K} & \multirow{2}{*}{$3054\pm446$} &  &  \\
     & & & if $V_{\rm scr\perp}$ \\ 
     \multirow{2}{*}{L} & \multirow{2}{*}{$5706\pm704$} &  & $ = 0$ km/s \\
     & & & \\ 
    \end{tabular}
    \caption{Scintillation arc curvatures and fractional screen distances $s$ at 1.4 GHz for B1929$+$10. Due to the pulsar's large transverse velocity \citep{chatterjee2004}, single solutions for $s$ were obtained assuming $V_{\rm psr\perp} \gg (V_{\rm obs\perp},V_{\rm scr\perp})$. Since $\psi$ was largely unconstrained by the observations, we show values of $s$ that would be obtained for characteristic values of $\psi$ noted in the righthand column. Features H--L have curvatures greater than the maximum possible values for $\psi = 0^\circ, V_{\rm scr\perp} = 0$ km/s, implying that either greater $\psi$ and/or greater $V_{\rm scr\perp}$ are required for these high-curvature arcs. Features J--L would require $\psi \gtrsim60^\circ$ for $V_{\rm scr\perp} = 0$ km/s; however, it is unclear whether these arcs really trace distinct screens (see main text).}
    \label{tab:B1929results}
\end{table}

\subsection{B1957$+$20}

A single, very weak and diffuse scintillation arc is detected for B1957$+$20 at curvature $\eta \geq 220$ m$^{-1}$ mHz$^{-2}$ (1.4 GHz). While the scintles in the dynamic spectrum do appear to be resolved in both frequency and time (see Figure~\ref{fig:all_spec2}), the low S/N of the pulsar required a longer integration time in the dynamic spectrum than for the other pulsars, yielding reduced resolution in the secondary spectrum. Due to the pulsar's large transverse velocity ($V_{\rm psr\perp} = 366^{+62}_{-98}$ km/s; \citealt{romani2022}), a single screen distance is estimated from the arc curvature to be $s \leq 0.44$ ($\psi \geq 0^\circ$), which corresponds to a physical screen distance $\gtrsim 1.5$ kpc from the observer. This pulsar is the only source in the dataset with an H$\alpha$-emitting bow shock, the stand-off radius of which is $\approx 7700 - 9300$ au, depending on the shock thickness \citep{romani2022}. The arc curvature that is inferred is far too large to be connected to the pulsar bow shock. 

B1957$+$20 is a well-studied black widow pulsar that exhibits strong plasma lensing near eclipse by its companion \citep{main2018,bai2022,lin2023}. Our observations were several hours away from eclipse, and do not display evidence of any scattering through the pulsar's local environment, whether that be the companion outflow or the pulsar bow shock. Previous observations away from eclipse measured a scattering timescale of $12$ $\mu$s at 327 MHz (equivalent to $\approx 0.04$ $\mu$s at 1.4 GHz or a scintillation bandwidth of $\approx 4$ MHz; \citealt{main2017}). Our observations imply a scintillation bandwidth $\Delta \nu_{\rm d}\approx 10$ MHz, based on fitting a 1D Lorentzian to the autocorrelation function of the dynamic spectrum, which is larger than the equivalent $\Delta \nu_{\rm d}$ for \cite{main2017}.

\section{Constraints on Scattering Screens}\label{sec:screen_constraints}

Scintillation arc properties can be translated into physical constraints on the scattering medium. In the following sections, we consider possible interpretations of the scintillation arcs in our sample, including the relationship between their intensity distributions and interstellar density fluctuations (Section~\ref{sec:power_distributions}) and potential associations between arcs and pulsar bow shocks (Section~\ref{sec:bow_shock_screens}). In Section~\ref{sec:ism3d}, we contextualize the scattering media in relation to the larger-scale ISM by utilizing 3D models of discrete structures identified in continuum maps.

\subsection{Power Distribution in the Secondary Spectrum}\label{sec:power_distributions}

Many secondary spectra contain a bright core of power near the origin that is up to $\sim10^5$ times brighter than the power distributed along a scintillation arc. In our data set, this feature is most prominent for B0355$+$54, B0919$+$06, J1713$+$0747, and B1929$+$10. This bright central core can be interpreted as individual, weakly deviated ray paths ($\pmb{\theta}_1 = \pmb{\theta}_2$ in Eq.~\ref{eq:ft}), whereas arcs are sensitive to lower intensity radiation that can trace a much larger extent of the scattered disk than other scattering measurements (e.g., scintillation bandwidths or pulse broadening times). 

Arc properties can be evaluated in the context of weak and strong scintillation, which correspond to the regimes where the modulation index (rms intensity variation / mean intensity) is $m\ll1$ and $m \sim 1$, respectively. In our dataset, B0950$+$08 and B1929$+$10 are both weakly scintillating ($m\approx 0.1$ and $m\approx 0.2$, respectively), whereas all of the other pulsars have $0.7 \lesssim m \lesssim 1$. Multiple, high-contrast (thin) arcs are usually detected in weak scintillation because there is still significant undeviated radiation incident on each scattering screen; typically this regime applies to lower DM pulsars, as seen here for B1929$+$10. Higher DM pulsars often fall in the strong scintillation regime, where arcs tend to be more diffuse and lower contrast \citep{stinebring2022}, as seen for B0355$+$54 and J1643$-$1224 (for which we find $m \approx 0.9$ and $m\approx1$, respectively). However, this trend does not appear to be clear cut; e.g., J1713$+$0747 displays a thin, highly defined arc despite having $m \approx 0.7$ in our dataset. Given that scattering is highly chromatic, arc properties also tend to evolve with frequency \citep[e.g.][]{stinebring2019}, and many scintillation arcs detected at low ($<500$ MHz) frequencies appear to be thicker and lower contrast \citep{wu2022,stinebring2022}. We therefore expect that the strongly scintillating pulsars in our dataset, such as B0355$+$54, J1643$-$1224, and J1740$+$1000, could yield multiple additional arcs if observed at higher frequencies.

\subsubsection{Relevance of Power-Law Electron Density Fluctuations}

In the limit of weak scintillation, the power distribution along an arc can be derived for a density fluctuation wavenumber spectrum of index $\beta$ to be $S_2(f_\lambda)\propto f_\lambda^{-(\beta+1)/2}$, or $S_2(f_\lambda)\propto f_\lambda^{-7/3}$ for $\beta = 11/3$, a Kolmogorov spectrum (\citealt{cordes2006}, Appendix D; \citealt{reardon2020}). In full, $S_2(f_t,f_\lambda)$ also includes a constant factor that depends on the screen distance $s$, the transverse velocity $V_\perp$, and a resolution function that accounts for the effect of finite sampling of the dynamic spectrum. 

\begin{figure*}
    \centering
    \includegraphics[width=\textwidth]{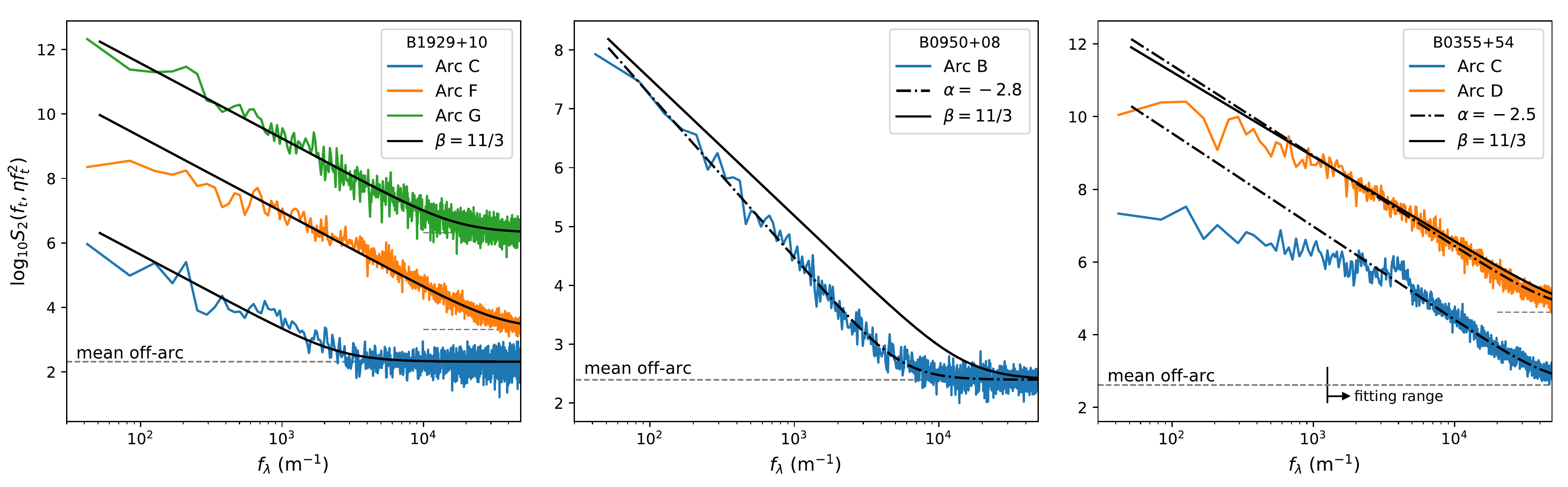}
    \caption{Arc intensity as a function of $f_\lambda$ for B1929$+$10 (left), B0950$+$08 (middle), and B0355$+$54 (right). For B1929$+$10, arcs C (blue), F (orange), and G (green) are shown, and arcs F and G are purposely offset from the mean off-arc noise (grey dashed line). Arcs C (blue) and D (orange) are shown for B0355$+$54, with an arbitrary vertical offset for arc C. The black curves in each panel show power-law models for the arc intensity, with black solid curves corresponding to $\alpha = -(\beta+1)/2 = -7/3$, where $\beta = 11/3$ is the expectation for Kolmogorov density fluctuations. For B1929$+$10, the best-fit spectral indices for all three arcs are consistent to within $1\sigma$ with $\alpha = -7/3$. For B0950$+$08 and B0355$+$54, the black dashed curves show the best-fit power laws with spectral indices $\alpha$ shown in the legends. Neither arc for B0355$+$54 is consistent with a single power-law intensity distribution, and the power laws shown were fit using $f_\lambda > 10^3$ m$^{-1}$. }
    \label{fig:powerlaw_fits}
\end{figure*}

To assess whether arcs are broadly consistent with arising from a power-law wavenumber spectrum of electron density fluctuations, we examine the distribution of power within the brightest scintillation arcs as a function of $f_\lambda$. We examine four pulsars: B1929$+$10, B0950$+$08, B0355$+$54, and J1643$-$1224. Of these, two are weakly scintillating (B1929$+$10, B0950$+$08) and two are strongly scintillating (B0355$+$54, J1643$-$1224) based on the modulation indices of their dynamic spectra. The power contained within each arc was summed along the $f_\lambda$ axis and fit as a power-law with two free parameters, an amplitude and a spectral index $\alpha = -(\beta+1)/2$, where $\beta$ is the index of the electron density fluctuation spectrum. Figure~\ref{fig:powerlaw_fits} shows the arc power compared to the best-fit models for B1929$+$10, B0950$+$08, and B0355$+$54.  

For B1929$+$10, all three arcs examined (arcs C, F, and G, which had the most precise curvature measurements) have best-fit spectral indices consistent with $\alpha = -7/3$, the expectation for Kolmogorov density fluctuations. The brightest arc examined, arc F, deviates from the power-law at low $f_\lambda$. B0950$+$08 yields a best-fit power-law index $\alpha=-2.8\pm0.1$ for arc B, which corresponds to $\beta > 4$ and may be indicative of refraction \citep{goodman85}. For B0355$+$54, we find $\alpha=-2.5\pm0.1$ for both arcs C and D, although arc C has discrete clumps that deviate significantly from a uniform power-law intensity distribution. Both arcs C and D for B0355$+$54 also show roll-offs at low $f_\lambda$, similar to arc F for B1929$+$10. While sampling near the origin of the secondary spectrum is more limited, it is possible that these roll-offs are related to multi-scale structure in the scattered image. Interestingly, the arc intensity distributions for B0355$+$54 are consistent to within $2\sigma$ from a Kolmogorov power-law at large $f_\lambda$. We also investigate the power distribution for J1643$-$1224 and find  $\alpha \approx -1.6$, a large departure from the Kolmogorov expectation that could partially be due to our limited resolution of the arc's full extent in $f_\lambda$ (see Section~\ref{sec:J1643}). 

One possible interpretation of these arc intensity distributions is that they have been modified from a Kolmogorov form by some combination of astrophysical and instrumental effects. Here we consider the potential relevance of three main effects, following Section 5.2 of \cite{cordes2006}:
\begin{enumerate}
    \item \textit{Inner scale:} Arcs are truncated when the diffraction spatial scale $l_{\rm d}$ becomes comparable to the inner scale $l_{\rm i}$ of the density wavenumber spectrum. For a 1D scattering angle $\theta_{\rm d}$, the diffraction scale is $l_{\rm d} \sim (\theta_{\rm d} k)^{-1}$ for a wavenumber $k$. For a scattering time $\tau_{\rm d} \sim \theta_{\rm d}^2/c$, the diffraction scale is then 
    \begin{align}
        l_{\rm d} &\approx \frac{1}{2\pi\nu}\bigg[\frac{c}{\tau_{\rm d}}\bigg(\frac{d_{\rm so}d_{\rm lo}}{d_{\rm sl}}\bigg)\bigg]^{1/2} \\
        &\approx \frac{1.5\times10^4\ {\rm km}}{\nu_{\rm GHz}}\bigg[\frac{1}{\tau_{\rm d,\mu s}}\bigg(\frac{d_{\rm so}d_{\rm lo}}{d_{\rm sl}}\bigg)_{\rm kpc}\bigg]^{1/2},
    \end{align}
    where $d_{\rm so}$, $d_{\rm sl}$, and $d_{\rm lo}$ are the source-observer, source-lens, and lens-observer distances. For a single screen, the maximum arc extent in $f_t$ due to this effect is approximately $f_{t,\rm inner} \sim (V_\perp/\lambda s) q_{\rm i} l_d$, where $q_{\rm i} = 2\pi/l_{\rm i}$. For B0950$+$08, $s\approx0.05$ and $\tau_{\rm d} < 1\ \mu \rm s$, implying $l_{\rm d} \gtrsim 2\times10^4$ km. We thus find that $l_i$ would have to be implausibly large, given typical inferred values $<1000$ km \citep{spangler1990,armstrong95,bhat2004,rickett2009} to explain the steep drop-off in arc power for B0950$+$08. 
    For B1929$+$10, taking nominal screen parameters for arc F ($s \approx 0.4$) implies $l_{\rm d} \approx 3000$ km, which places an upper limit on the inner scale that is consistent with other inferred values.
    \item \textit{Finite source size and multiple screens:} Arc extent depends on the angular scale of coherent radiation incident on the scattering screen, which we denote $\theta_{\rm scr}$. This angular scale is determined by both the finite size of the pulsar emission region and any scattering through additional screens. Scintillations will be quenched when the coherence length of radiation incident on the scattering screen, $l_{\rm c} \approx \lambda / 2\pi \theta_{\rm scr}$, is of order the size of the scattering cone at the screen, $l_{\rm cone} \approx \theta_{\rm obs}d_{\rm lo}$, where $\theta_{\rm obs} = s\theta_{\rm scr}$. In the simplest (single screen) case, arcs will be suppressed beyond $f_{t, \rm sou} = V_\perp[D(1-s)\theta_{\rm obs}]^{-1}$ \citep{cordes2006}. A screen close to the source could have larger $\theta_{\rm scr}$ yielding smaller $f_{t, \rm sou}$, if additional screens are not present. On the other hand, scattering through one screen can reduce $l_{\rm c}$ for a subsequent screen, which could in principle lead to weaker, more truncated arcs for larger screen distances $s$. Of the pulsars considered here, this effect is most likely relevant to B0355$+$54, as the pulsar is in strong scintillation and hence more likely to have significant scattered radiation incident on each of the four screens along the LOS. 
    
    First, we consider the possibility that the shallowest arc corresponds to a screen close to the pulsar, and examine whether the finite size of the pulsar emission region could affect the arc extent (ignoring, for now, the presence of additional screens). For an emission region size $\sim 100$ km and a screen $\lesssim 1$ pc from the pulsar, $f_{t, \rm sou} \sim 1$ Hz, orders of magnitude greater than the observed extent of the arc. To explain the observed arc extent, the screen would need to be $\lesssim 100$ au from the pulsar, far smaller than the estimated bow shock stand-off radius of $\sim 8000$ au. Next, we consider the possibility that scattering through multiple screens modifies the intensity distributions of the brightest arcs for B0355$+$54, arcs C and D (Figure~\ref{fig:powerlaw_fits}). While both arcs are broadly consistent with the same power-law, $\alpha = -2.5\pm0.1$, arc C shows a stronger deviation at lower $f_\lambda$. Unfortunately, the twofold ambiguity in screen location means that it is unclear which order the screens are encountered; i.e., arc C could be produced prior to arc D, or after. However, both arcs extend across the full Nyquist range in $f_\lambda$ and have similar amplitudes of intensity, suggesting that neither arc is significantly suppressed by the presence of a preceding screen. 
    \item \textit{Sensitivity limitations:} If an arc is low intensity and/or poorly resolved in the secondary spectrum, then it can appear to be truncated because its power-law drop-off makes it indistinguishable from the noise at smaller ($f_t$, $f_\lambda$) than for a higher-intensity arc. This effect is likely most relevant to the shallowest arcs detected for B0355$+$54, B0919$+$06, B0950$+$08, and B1929$+$10. 
\end{enumerate}

These findings suggest that while B1929$+$10 has scintillation arc intensities consistent with diffractive scintillation produced by a turbulent density fluctuation cascade, B0950$+$08 is likely affected by additional refraction. Similarly, the discrete clumps of power in arc C for B0355$+$54, coupled with the significant roll-off in arc intensity at small $f_\lambda$, suggest non-uniform, multi-scale structure in the scattered image that is also produced by refraction. Overall, these features can be interpreted as resulting from a superposition of refracting plasma structures (blobs or sheets) and the nascent density fluctuations associated with interstellar turbulence.

\subsection{Near-Pulsar Screens \& Candidate Bow Shocks}\label{sec:bow_shock_screens}

Three pulsars in our sample have low-curvature arcs that could arise within the pulsars' local environments, B0355$+$54, B0950$+$08, and B1929$+$10. B0950$+$08 does not have a directly imaged PWN or bow shock, although recently \cite{ruan2020} have argued that off-pulse radio emission detected up to $\sim 100^{\prime\prime}$ from the pulsar location is consistent with arising from a PWN. Both B0355$+$54 and B1929$+$10 have ram pressure confined PWNe identified in X-ray and radio, and are likely to have bow shocks \citep{wang93,becker2006,mcgowan2006}. In the case of B0355$+$54, we have acquired optical H$\alpha$ imaging data from Kitt Peak National Observatory (KPNO) with the Nicholas U. Mayall 4-meter Telescope on 25-Oct 2017, using the Mosaic-3 detector. The observations were part of a larger campaign to search for H$\alpha$ bow-shocks which are a publication in preparation. The target list included B0355$+$54 for 600\,s, deeper available data than from the INT/WFC Photometric H-Alpha Survey \citep{2014MNRAS.444.3230B}. On the same night, we observed the Guitar Nebula for the same amount of time at the same detector location. No bow-shock structure was observed for B0355$+$54. By fractionally adding the Guitar Nebula image to the sky background nearby until it became visible and then decreasing the fraction by increments of 0.05 until the known bow shock faded into the background, we estimate a non-detection limit of about $15\%$ of the Guitar Nebula apex flux. In units of H$\alpha$ photons, any bow-shock from B0355$+$54 would therefore have an apex surface brightness flux of $\lesssim 5.4\times10^{-4}$ $\gamma/$cm$^{2}$s$^{-1}$, using the known flux of the Guitar Nebula from \citet{brownsberger2014}.

We now consider the range of screen conditions ($\psi, V_{\rm scr\perp}$) that would be needed for the low-curvature arcs to be associated with these pulsars' bow shocks, if the bow shocks exist.

For B0355$+$54, the lowest arc curvature is $\eta = 2.7\pm1.2$ m$^{-1}$ mHz$^{-2}$. For the pulsar's measured spin-down luminosity and transverse velocity (Table~\ref{tab:source_list}), we estimate a bow shock stand-off radius $R_0 \approx 7900 - 24000$ au for electron densities $\sim 0.1 - 0.01$ cm$^{-3}$ (Equation~\ref{eq:r0}). Given the parallax distance of $D = 1.09^{+0.23}_{-0.16}$ kpc \citep{chatterjee2004}, we thus estimate an upper limit on the fractional screen distance $s \approx R_0/D \sim 9\times10^{-5}$ for $R_0 \approx 24000$ au and $D = (1.09-0.16)$ kpc. Given that the bow shock nose is likely inclined relative to the LOS, $s$ could be even larger. The measured arc curvature can accommodate $s \sim 9\times10^{-5}$ for $\psi \approx 50^\circ$ leaving $V_{\rm scr\perp}$ small, or alternatively, for small $\psi$ ($<45^\circ$) and $V_{\rm scr\perp} \gtrsim 20$ km/s. Previous studies have inferred screen velocities ranging up to tens of km/s and similarly wide ranges of screen angles \citep[e.g.][]{reardon2020,mckee2022}. We thus conclude that the lowest curvature arc for B0355$+$54 could be consistent with a scattering screen at the bow shock, but more observations are needed to determine whether the screen is indeed close to the pulsar or close to the observer.

For B0950$+$08, the lowest arc curvature is $\eta = 17\pm1$ m$^{-1}$ mHz$^{-2}$, and the nominal stand-off radius ranges from $R_0 \approx 1480 - 4440$ au for $n_e \approx 0.1 - 0.01$ cm$^{-3}$. Following a similar line of reasoning as for B0355$+$54, we find that the lowest curvature arc can be consistent with a scattering screen at the bow shock if $\psi \gtrsim 70^\circ$ and $V_{\rm scr\perp} \sim 20$ km/s. These constraints can be relaxed somewhat if the ISM density is even lower and/or if the shock widens significantly where it is intersected by the pulsar LOS.

For B1929$+$10 the lowest arc curvature is $\eta = 4.0\pm0.7$ m$^{-1}$ mHz$^{-2}$ and the nominal stand-off radius ranges from $R_0 \approx 805 - 2400$ au for $n_e \approx 0.1 - 0.01$ cm$^{-3}$, or $s\sim10^{-5}$. In this case, we find that even if the shock is widened significantly at the pulsar LOS, large values of $\psi$ and $V_{\rm scr\perp}$ are still needed to bring the screen distance into agreement with the measured arc curvature (see Figure~\ref{fig:B1929_zoom}). E.g., assuming the shock corresponds to a screen distance $s\sim10^{-4}$ (equivalent to a distance of about 7500 au from the pulsar), the arc curvature would imply $\psi > 75^\circ$ and $V_{\rm scr\perp} > 100$ km/s. 

While all three pulsars could have arcs associated with their putative bow shocks, both B0950$+$08 and B1929$+$10 require more restricted ranges of $\psi$ and $V_{\rm scr\perp}$ in order for the screen distance to be broadly consistent with the bow shock. Nonetheless, there is a considerable range of ISM densities and shock inclination angles that are possible. If future observations are able to constrain the arc curvatures over time and determine that these arcs are from the pulsars' sub-parsec environments, then the resulting screen constraints could be used to infer the inclination angles of the bow shocks, the radial velocity components of the pulsars, and a more restricted range of local ISM densities. 

\subsection{Associations with Foreground Structures}\label{sec:ism3d}
\begin{figure}
    \centering
    \includegraphics[width=0.45\textwidth]{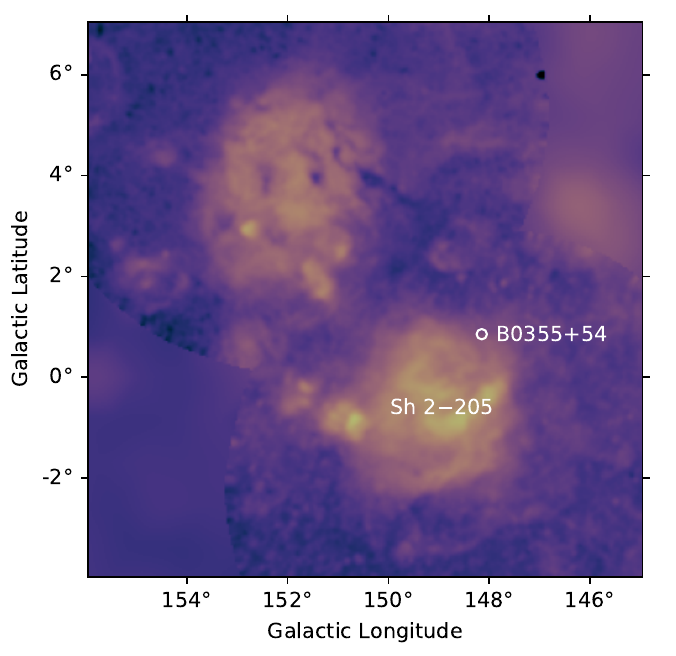}
    \caption{H$\alpha$ emission observed in a $10^\circ$ by $10^\circ$ area around the LOS to B0355$+$54 (from the all-sky map provided by \citealt{finkbeiner2003}). The pulsar LOS intersects the edge of an HII region included in the Sharpless catalog, Sh $2-205$ \citep{sharpless1959}, and may plausibly account for one of the scintillation arcs observed from this pulsar. }
    \label{fig:sh205}
\end{figure}

A search for associations between each pulsar LOS and foreground continuum sources catalogued in the Simbad database recovered known associations for several pulsars, including the HII region Sh $2-27$ for J1643$-$1224 \citep{harveysmith2011} and the HII region Sh $2-205$ for B0355$+$54 \citep{mitra2003}. As shown in Figure~\ref{fig:sh205}, B0355$+$54 intersects the edge of Sh $2-205$, which is approximately 1 kpc away and 24 pc in diameter \citep{romero2008}.
While there is twofold ambiguity in the screen distances inferred for B0355$+$54, one of the near-pulsar screen solutions does coincide with the HII region. We also find a new potential association for the LOS to B1957$+$20, which passes within $1.4^{\prime \prime}$ of a star in Gaia DR3 (ID:1823773960079217024) that has a parallax of $0.7\pm0.5$ mas \citep{gaia_catalog}, which is not the white dwarf companion of the pulsar (Gaia ID: 1823773960079216896). Given the large uncertainties on the foreground star's parallax, it is unclear whether the star is intersected by the pulsar LOS; however, the nominal screen distance inferred from the arc curvature for B1957$+$20 is $1.5$ kpc, somewhat similar to the nominal distance of the star, 1.4 kpc. No other novel associations were found in Simbad for the pulsars in the sample. 

\begin{figure*}
    \centering
    \includegraphics[width=0.9\textwidth]{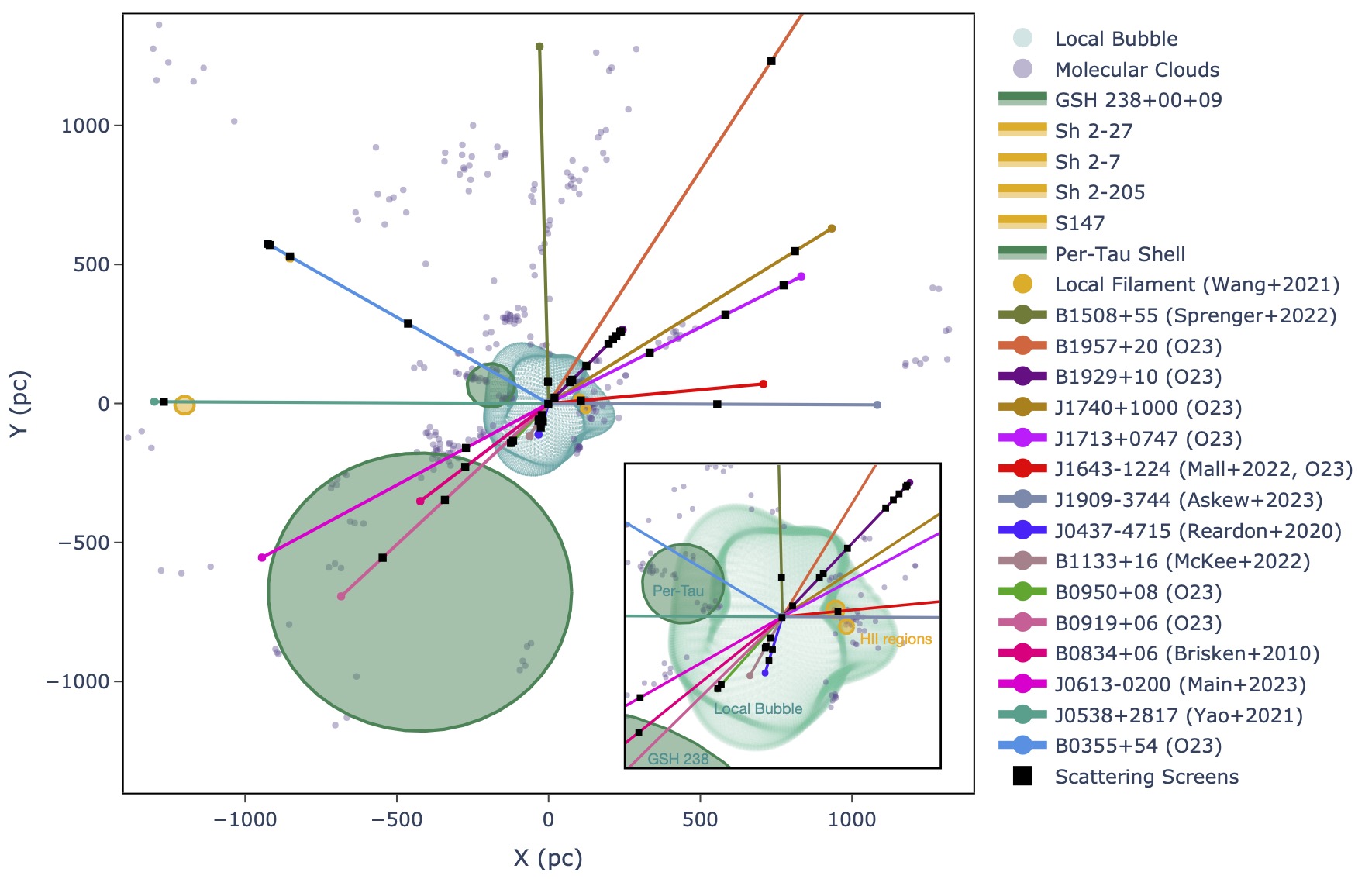}
    \caption{Locations of pulsar LOSs, scattering screens inferred from scintillation arcs, and simple models of discrete ISM features based on continuum maps, in heliocentric Galactic Cartesian coordinates looking down onto the Galactic plane. An inset shows a close-up of the region $\pm300$ pc around the origin. A total of 14 pulsars are shown, eight from this work (O23) and seven from previous studies noted in the legend. Pulsar names in the legend are ordered clockwise starting from the LOS to B1508$+$55, which is located nearly parallel to the $Y$-axis at $X=0$ pc. {Scattering screens shown for pulsars from this work correspond to the near-pulsar solutions for $\psi = 0^\circ$ and $V_{\rm scr\perp} \ll V_{\rm obs\perp}, V_{\rm psr\perp}$. The near-pulsar solutions are favored for pulsars with large transverse velocities (B0919$+$06, J1740$+$1000, B1929$+$10, and B1957$+$20), whereas for pulsars with lower transverse velocities (B0355$+$54, B0950$+$08, J1643$-$1224, and J1713$+$0747) the two-screen solution cannot be formally broken by our observations. These screen distances are thus notional and have formal uncertainties largely dominated by the uncertainties on the arc curvature (see Figure~\ref{fig:screen_examples_with_errors} for examples).} The screens shown from other pulsar studies are more precisely determined from either arc curvature variations or VLBI scintillometry. Models for discrete ISM features include the Local Bubble, based on a spherical harmonic decomposition of dust extinction boundaries (here we show the decomposition mode $l = 6$ from \citealt{pelgrims2020}), the superbubble GSH 238$+$00$+$09 \citep{heiles98,lallement2014}, and the Per-Tau Shell \citep{bialy2021}. Local molecular clouds are also shown \citep{zucker2020}. Three HII regions (Sh 2$-$7, 2$-$27, and 2$-$205) and one supernova remnant (S147, associated with pulsar J0538$+$2817) are shown. The spatial parameters used to model each ISM feature are explained in Appendix~\ref{app:ISM}. A 3D interactive version of this figure is available at \url{https://stella-ocker.github.io/scattering\_ism3d\_ocker2023}. The interactive version can be zoomed, rotated, and modified to only show specific legend entries.}
    \label{fig:ism3d}
\end{figure*}

The boundary of the Local Bubble has long been attributed a role in pulsar scattering \citep[e.g.][]{bhat1998}. Recent studies have leveraged Gaia to map dust extinction and molecular clouds demarcating the edge of the Bubble in exquisite detail \citep{lallement2019,pelgrims2020,zucker2022}, in addition to revealing large-scale structures such as the Radcliffe Wave \citep{alves2020}, the ``Split'' \citep{lallement2019}, and the Per-Tau Shell \citep{bialy2021}. In Figure~\ref{fig:ism3d} we compare the pulsar LOSs in our sample to modeled foreground structures, including the inner surface of the Local Bubble \citep{pelgrims2020}, the superbubble GSH 238$+$00$+$09 \citep{heiles98,lallement2014}, the Per-Tau Shell \citep{bialy2021}, and several HII regions confirmed to intersect pulsar LOSs \citep{mitra2003,harveysmith2011,ocker2020,mall2022}. 
We have also included all of the local molecular clouds catalogued by \cite{zucker2020}, which trace the large-scale structure of the Radcliffe Wave and the Split. While the molecular clouds themselves are not expected to induce scattering, electron density enhancements in the partially ionized gas surrounding these clouds are, in theory, potential locations of enhanced scattering. The spatial parameters used to model each ISM feature are explained in Appendix~\ref{app:ISM}.

\begin{figure*}
\centering
    \includegraphics[width=0.9\textwidth]{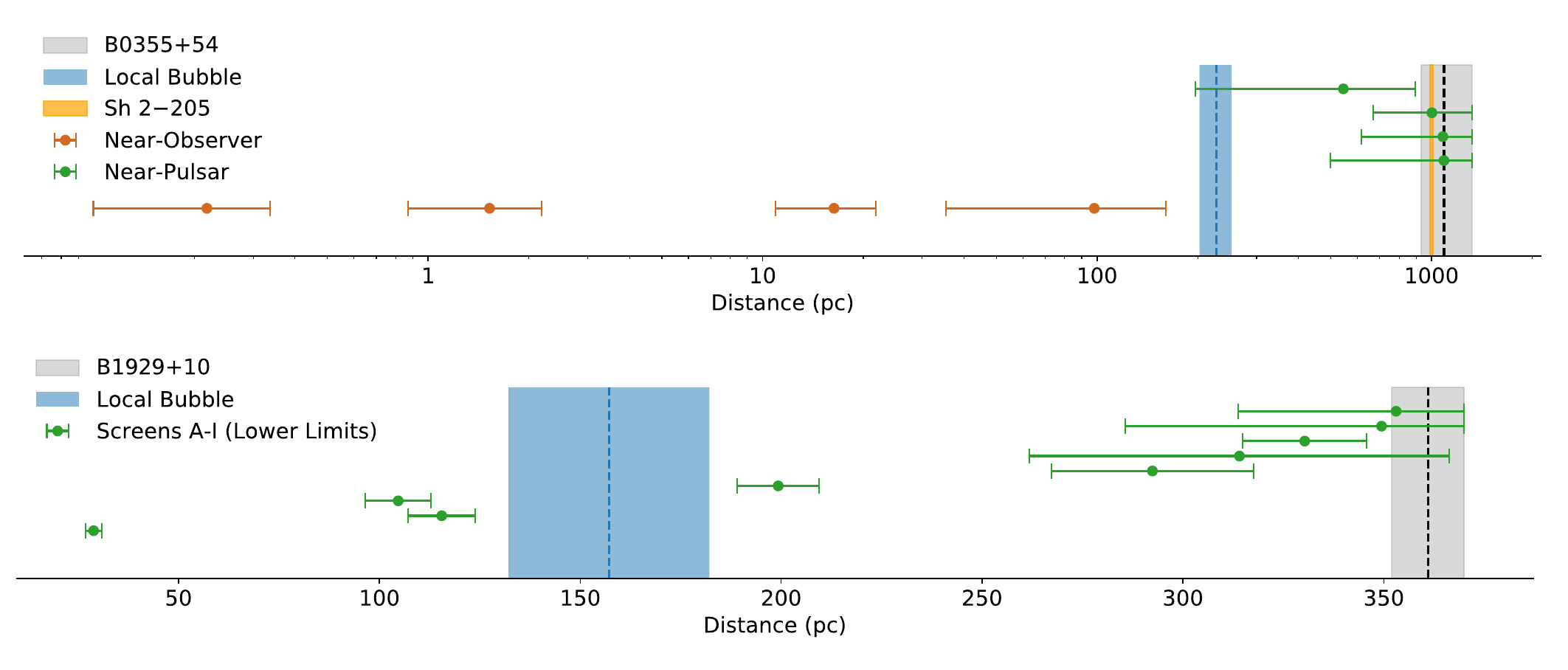}
    \caption{Screen distance estimates for B0355$+$54 (top) and B1929$+$10 (bottom). The pulsar locations at $1.09^{+0.23}_{-0.16}$ kpc (B0355$+$54) and $361\pm9$ pc (B1929$+$10) are shown by the black dashed lines and shaded grey error regions. Screen locations are shown with an arbitrary vertical offset for visualization purposes. The errors shown on the screen distances include the uncertainties in pulsar distance, transverse velocity, and scintillation arc curvature, but do not account for the unconstrained $\psi$ and $V_{\rm scr\perp}$. In all of these cases, the error bars shown are largely dominated by the uncertainty in arc curvature. For B0355$+$54 there is degeneracy between the near-pulsar solutions (green points) for screen distance and the near-observer solutions (brown points). The near-pulsar screen distances should be regarded as lower limits, whereas the near-observer distances are upper limits. B1929$+$10 has a large enough transverse velocity to yield single distance estimates for each screen, and for this pulsar only screens for Arcs A--I are shown; these screen distances are lower limits. The blue dashed lines indicate where each LOS crosses the inner surface of the Local Bubble, and the shaded blue regions indicate $\pm25$ pc around the intersection point. For B0355$+$54, the adopted location and width of the HII region Sh 2$-$205 are shown in yellow.}
    \label{fig:screen_examples_with_errors}
\end{figure*}

Figure~\ref{fig:ism3d} shows the locations of scattering screens inferred from scintillation arcs. For simplicity, the screen locations are shown as point estimates for $\psi = 0^\circ$ and only the near-pulsar solutions where relevant; these screen distance estimates thus have substantial uncertainties and are only notional. Formally, the uncertainties on the screen distance estimates are often dominated by the uncertainties of the arc curvatures, as all of the pulsars (barring J1740$+$1000, which has no parallax) have fractional distance and transverse velocity uncertainties $\lesssim 20\%$. {For pulsars with large transverse velocities (B0919$+$06, J1740$+$1000, B1929$+$10, and B1957$+$20), the screen distance estimates shown in Figure~\ref{fig:ism3d} correspond to lower limits, and any of these screens could be closer to the pulsar for larger  $\psi$ or $V_{\rm scr\perp}$. For pulsars with low transverse velocities (B0355$+$54, B0950$+$08, J1643$-$1224, and J1713$+$0747) the uncertainties on the screen distance estimates shown in Figure~\ref{fig:ism3d} are even less constrained, as there are both near-pulsar and near-observer solutions each with unknown $\psi$ and $V_{\rm scr\perp}$. Examples of the screen distance uncertainties for B0355$+$54 and B1929$+$10 are shown in Figure~\ref{fig:screen_examples_with_errors}. Despite these uncertainties, we are able to make some initial comparisons to known ISM features below, which highlight LOSs of interest for future study.}

More precise screen locations are also shown in Figure~\ref{fig:ism3d} for seven additional pulsars with scintillation arcs that are well-characterized in previous works: J0437$-$4715 \citep{reardon2020}, J0538$+$2817 \citep{yao2021}, J0613$-$0200 \citep{main2023b}, B0834$+$06 \citep{brisken2010}, B1133$+$16 \citep{mckee2022}, B1508$+$55 \citep{sprenger2022}, and J1909$-$3744 \citep{askew2023}. These pulsars were selected from the literature because their scintillation properties were characterized to high precision using either arc curvature variations or VLBI scintillometry, but in future work we will expand our analysis to a broader sample. Readers are strongly encouraged to view a 3D interactive version of the figure that has been optimized for the complexity of the data.\footnote{\url{https://stella-ocker.github.io/scattering\_ism3d\_ocker2023}} 

Several of the pulsars shown in Figure~\ref{fig:ism3d} have scattering screens well within the dust boundary of the Local Bubble, including J0437$-$4715, B1133$+$16, J1643$-$1224, B1508$+$55, and B1929$+$10. Of these, B1133$+$16 and B1929$+$10 both have screens within 30 pc of the Sun, which could lie near or within local interstellar clouds \citep{frish2011,linsky2022}. B0355$+$54, B0950$+$08, and J1713$+$0747 could also have screens associated with the local interstellar clouds, if follow-up observations resolve their twofold screen location ambiguities. Pulsars B0919$+$06, B0834$+$06, and J0613$-$0200 all have LOSs near the superbubble GSH 238$+$00$+$09, with J0613$-$0200 actually intersecting the bubble for as much as 500 pc. This superbubble may extend farther above the Galactic plane (higher $Z$) than the rough representation in the 3D version of Figure~\ref{fig:ism3d} \citep{ocker2020}. One pulsar LOS in Figure~\ref{fig:ism3d} directly intersects a cluster of local molecular clouds, but shows no evidence of scattering accrued by the intersection: J1909$-$3744 passes through Corona Australis at about 150 pc from the observer, but shows evidence for only one dominant scattering screen at a distance of about 600 pc \citep{askew2023}.  

It remains difficult to associate any of the scattering screens presented here with the boundary of the Bubble, due not only to uncertainties in the scattering screen distances but also the modeled Bubble surface.
The Local Bubble surface shown in Figure~\ref{fig:ism3d} represents the inner surface of the Bubble (not the peak extinction), which could be offset from any related ionized scattering structure by as much as 25 pc or more. The exact offset expected between the inner surface of the Bubble traced by dust and any plasma boundary relevant to radio scattering is difficult to estimate, as it depends on the 3D distribution of stars, their parallax uncertainties, the uncertainties on individual extinction to the stars, and the specifics of the inversion algorithms used to infer the dust extinction boundary. Recently, \cite{liu2023L} argued that scattering screens for J0613$-$0200 and J0636$+$5128 are associated with the edge of the Local Bubble, based on the same dust extinction maps that informed the model used here \citep{lallement2019,pelgrims2020}. Given that the Bubble is such a large-scale feature, one would expect there to be evidence of scattering screens at the edge of the Bubble for many more pulsar LOSs, and it remains possible that follow-up observations of the pulsars studied here will reveal additional evidence connecting pulsar scintillation arcs to the Bubble's boundary. However, making such connections will require ruling out the possible chance coincidence of many small scattering structures, as our observations of B1929$+$10 indicate that scintillation arcs can evidently be produced in large numbers far from the Bubble surface.

\section{Discussion}\label{sec:conc}

\subsection{Key Results}
In this study we have conducted sensitive observations of scintillation arcs for eight pulsars using FAST. Scintillation arcs were detected from all pulsars in the study, tracing a broad distribution of scattering structures in the local ISM. Several pulsars in our sample show low-curvature, truncated arcs. For B0355$+$54, B0950$+$08, and B1929$+$10 these arcs could be associated with their putative bow shocks for a plausible range of screen configurations and ISM densities. Comparison of scattering screen constraints to local ISM structures observed in multi-wavelength continuum maps also suggests that one of the scattering screens for B0355$+$54 could coincide with the HII region Sh 2$-$205. Follow-up observations are needed to confirm or deny these associations.

At least nine arcs are observed toward B1929$+$10, which is just $361\pm9$ pc away \citep{chatterjee2004}. This finding demonstrates that with sufficient sensitivity, weakly scintillating, nearby pulsars can reveal a remarkably high concentration of scattering screens. {B1929$+$10 is also one of only a few pulsars that shows evidence of TSAS detected via time-variable HI absorption of the pulsar spectrum \citep{stanimirovic2010,zweibel_review18}.} The possible prevalence of arc ``forests'' (as seen for another pulsar by D. Reardon et al., submitted) illustrates a strong need for scintillation arc theory that can accommodate $\gg2$ screens. A high number density of arcs and screens for nearby pulsars may support a picture in which more distant, strongly scintillating pulsars trace an extended medium made up of many screens \citep[e.g.][]{stinebring2022}. However, it remains possible that highly specific conditions are needed to observe many arcs at once (e.g., some combination of observing conditions including radio frequency and sensitivity, and astrophysical conditions including screen strength and alignment). One possibility is that packed distributions of screens only occur in certain ISM conditions. For example, B0950$+$08, the other nearby, weakly scintillating pulsar in our sample, shows only two arcs and has an overall deficit of scattering compared to other pulsars at comparable distances, suggesting that its LOS may be largely dominated by the hot ionized gas thought to pervade the Local Bubble. These mixed findings imply a clear need for a uniform, deep census of scintillation arcs towards pulsars within 500 pc of the Sun, {ideally through a commensal study of both arcs and TSAS to elucidate the relationship between small-scale structure in both ionized and atomic phases of the ISM}. 

\subsection{Origins of Scattering Screens}
One of the core questions at the heart of scintillation arc studies is to what extent arcs are produced by scattering through nascent density fluctuations associated with extended ISM turbulence, or through non-turbulent density fluctuations associated with discrete structures. Both of these processes can produce arcs, albeit of different forms. The variety of arc properties seen even within our sample of just eight pulsars broadly affirms a picture in which pulsar scattering is produced through a mixture of turbulence and refractive structures whose relevance depends on LOS, and likely also observing frequency. Of the pulsars shown in Figure~\ref{fig:ism3d}, there are few direct and unambiguous connections between their scattering screens and larger-scale ISM features, even for those pulsars with precise scattering screen distances. To some degree this lack of association is to be expected, as scintillation traces ISM phenomena at much smaller spatial scales than typical telescope resolutions. In future work we will expand upon the local ISM features shown in Figure~\ref{fig:ism3d} to examine a larger census of potential scattering media (e.g., the Gum Nebula, known HII regions, etc.). 

The ISM contains a zoo of structures that are not always readily visible in imaging surveys and may not appear except in targeted searches. One example is stellar bow shocks, which can sustain turbulent wakes and emissive nebulae up to 1000s of au in scale, such as those seen for the H$\alpha$-emitting bow shock of B2224$+$65 \citep{1993Natur.362..133C} and the X-ray PWN of B1929$+$10 \citep{kim2020}. An updated Gaia census of stars within the solar neighborhood suggests a mean number density of stars $\sim 0.06-0.08$ pc$^{-3}$, depending on the stellar types included \citep{reyle2022}. Of these, only a fraction will have magnetosonic speeds fast enough to generate bow shocks (e.g. \cite{shull2023} assume a mean number density $\sim 0.01$ pc$^{-3}$ for stars with bow shocks). Bow shock nebulae $\sim 1000$ au in size will have a volume filling factor $f_V \approx N_{\rm bs}(R_{\rm bs}/R_{\rm ISM})^3 \sim 0.01(1000\ \rm au/1\ pc)^3 \sim 10^{-9}$ for a number of bow shocks $N_{\rm bs}$ with spatial extents $R_{\rm bs}$ within an ISM volume of radius $R_{\rm ISM}$. The equivalent mean free path is $\sim 1$ Mpc. {Allowing for larger bow shock sizes could bring the mean free path down to $\sim$~kpc.}
Regardless, this rough estimation suggests that bow shock nebulae could only comprise a very small fraction of scattering media along pulsar LOSs.

High-resolution magnetohydrodynamic simulations of thermally unstable turbulent gas suggest that dense, elongated plasmoids may be a ubiquitous feature of both the cold and warm phases of the ISM \citep{fielding2023}. These plasmoids have been simulated down to spatial scales $\sim 10^3$ au and can result in density deviations $\sim 10^3\times$ nominal values, in addition to changes in magnetic field direction across their current sheets. It thus appears possible that the extended ISM spontaneously produces some scattering structures through plasmoid instabilities, in addition to turbulence and deterministic processes involving stars and nebulae. Future work should compare the rate at which these plasmoids form, their lifetimes, and volume filling factor to the distribution of known scattering screens.

Pulsar scintillation remains one of the few astrophysical probes of sub-au to au-scale structures in the ISM. While the ubiquity of scintillation arcs is now well-established for many LOSs \citep{stinebring2022,wu2022,main2023}, high-resolution studies of pulsar scattered images using scintillometry have only been applied to a limited number of pulsars. While inferences of ISM structure at the spatial scales probed by scintillation will benefit greatly from application of scintillometry to a broader sample of LOSs, our study demonstrates that single-dish observations of scintillation arcs continue to provide insight, particularly as increasing telescope sensitivity and spectral resolution appears to reveal more arcs than previously identified for some pulsars. 

\section*{Acknowledgements}

SKO, JMC, and SC are supported in part by the National Aeronautics and Space Administration (NASA 80NSSC20K0784). SKO is supported by the Brinson Foundation through the Brinson Prize Fellowship Program. TD is supported by an NSF Astronomy and Astrophysics Grant (AAG) award number 2009468. VP acknowledges funding from a Marie Curie Action of the European Union (grant agreement No. 101107047). SKO, JMC, SC, DS, and TD are members of the NANOGrav Physics Frontiers Center, which is supported by NSF award PHY-2020265. The authors acknowledge the support staff at FAST for managing the observations used in this work, and David Pawelczyk and Bez Thomas at Cornell for their technical contributions to data transport and delivery. This work is based in part on observations at Kitt Peak National Observatory at NSF’s NOIRLab (NOIRLab Prop. ID 17B-0333; PI: T. Dolch), which is managed by the Association of Universities for Research in Astronomy (AURA) under a cooperative agreement with the National Science Foundation. The authors are honored to be permitted to conduct astronomical research on Iolkam Du’ag (Kitt Peak), a mountain with particular significance to the Tohono O’odham. TD and CG thank the Hillsdale College LAUREATES program and the Douglas R. Eisenstein Student Research Gift for research and travel support. This work also benefited from the input of Thankful Cromartie, Ross Jennings, Robert Main, Joseph Lazio, Joris Verbiest, Henry Lennington, Joseph Petullo, Parker Reed, and Nathan Sibert. This work made use of Astropy (http://www.astropy.org), a community-developed core Python package and an ecosystem of tools and resources for astronomy \citep{astropy2013, astropy2018, astropy2022}.

\section*{Data Availability}

Data is available upon request to the corresponding author (SKO), and unprocessed observations in filterbank format are available through the FAST Data Center by contacting fastdc@nao.cas.cn. The Python program and input data for creating the 3D version of Figure 12 are available at \url{https://github.com/stella-ocker/ism-viz}. The Local Bubble model is available on Harvard Dataverse: \url{https://doi.org/10.7910/DVN/RHPVNC}. KPNO data are available on the NOIRLab Astro Data Archive: \url{https://astroarchive.noirlab.edu/}. The molecular cloud distance catalog is available on Harvard Dataverse: \url{https://doi.org/10.7910/DVN/07L7YZ}. 

\bibliographystyle{mnras}
\bibliography{bib}

\appendix

\section{Application of the $\theta-\theta$ Transform to J1740$+$1000}\label{app:theta-theta}

\begin{figure}
    \centering
    \includegraphics[width=0.4\textwidth]{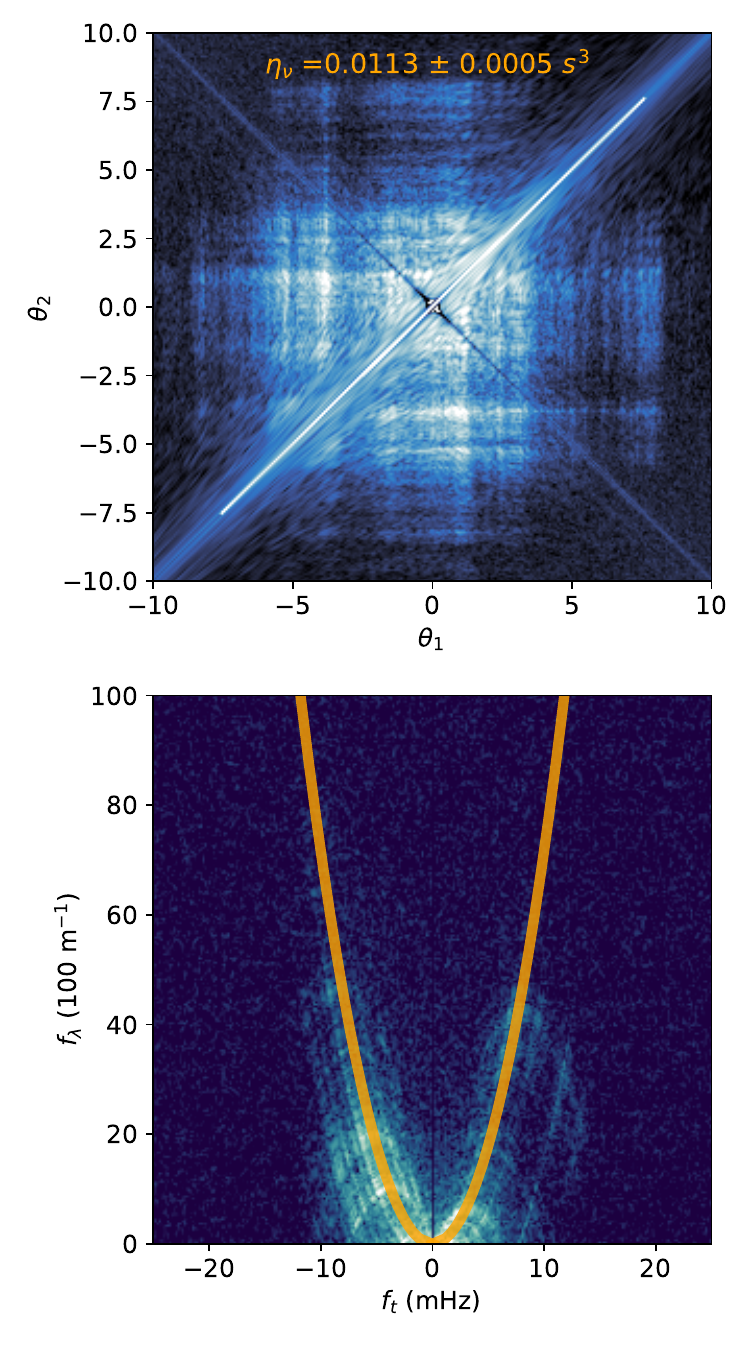}
    \caption{Top: The $\theta-\theta$ diagram for J1740$+$1000 evaluated for the best-fit curvature $\eta_\nu = 0.0113\pm0.0005$ s$^{3}$ at 1375 MHz, equivalent to $\eta = 72\pm5$ m$^{-1}$ mHz$^{-2}$. The main arc and reverse arclets are transformed into linear features that are parallel to the $\theta-\theta$ axes, as expected \citep{sprenger2021,baker2022}. Here $\theta_1$ and $\theta_2$ are scaled sky coordinates in units of fringe frequency $f_t$, and could in principle be transformed into angular sky coordinates given knowledge of the effective velocity $V_\perp$ \citep{baker2022}. Bottom: The secondary spectrum measured in the 1.4 GHz frequency band. The best-fit curvature inferred from the $\theta-\theta$ transform is shown as the orange curve, the width of which is equal to the measurement uncertainty.}
    \label{fig:theta-theta-J1740}
\end{figure}

The $\theta-\theta$ transform provides a highly precise method of measuring the curvatures of arcs that display well-defined reverse arclets \citep{sprenger2021}. This transform is applied here to J1740$+$1000, following the methodology laid out by \cite{sprenger2021} and \cite{baker2022}, and by making use of the \texttt{ththmod} module provided in the Python package \texttt{scintools} \citep{reardon2020}. The main premise of the transform is that by linearizing the secondary spectrum into $(f_t, f_\nu \cdot f_t^{-1})$ coordinates, the angular coordinates $(\theta_1, \theta_2)$ can be solved for assuming a single-screen, 1D scattered image (see Equations~\ref{eq:ft}-\ref{eq:fnu}). The result of the transform is the $\theta-\theta$ diagram, in which the power contained in the secondary spectrum is re-mapped onto a grid of $(\theta_1,\theta_2)$. When the curvature used to transform the secondary spectrum matches the curvature of the observed scintillation arc, the power contained within the arc is redistributed onto straight lines parallel to the $(\theta_1,\theta_2)$ axes. The arc curvature can thus be inferred by performing the $\theta-\theta$ transform over a range of possible curvatures, and finding the curvature that reproduces the expected distribution in $\theta-\theta$ space \citep{baker2022}. The $\theta-\theta$ diagram for J1740$+$1000 is shown in Figure~\ref{fig:theta-theta-J1740} for the best-fit curvature given by this method, $\eta_\nu = 0.0113\pm0.0005$ s$^{3}$ at 1375 MHz or equivalently $\eta = 72\pm5$ m$^{-1}$ mHz$^{-2}$. 

The $\theta-\theta$ diagram for J1740$+$1000 suggests that its reverse arclets are all consistent with a single curvature; that is, they arise from the same scattering screen. If the scattered image contained power from multiple screens (i.e. multiple curvatures) then there would be additional, distorted features in the $\theta-\theta$ diagram, as has been observed for B0834$+$06 \citep{baker2022b,baker2022}.

\section{Summary of Local ISM Features}\label{app:ISM}

\begin{table*}
    \centering
    \begin{tabular}{c | c | c | c}
    Feature & Center (pc) & Radius (pc) & References \\ 
            & $(X,Y,Z)$ & $(R_X,R_Y,R_Z)$ & \\ \hline \hline
    Local Bubble & see text & see text & \cite{pelgrims2020} \\
    GSH 238$+$00$+$09 & $(-424,-678,-302)$ & $(500,500,300)$ & \cite{heiles98,lallement2014,puspitarini2014} \\
    Per-Tau Shell & $(-190,65,-84)$ & $78$ & \cite{bialy2021} \\
    Molecular Clouds & see text & see text & \cite{zucker2020} \\
    Sh 2$-$7 & $(123,-19,55)$ & $14$ & \cite{finkbeiner2003,miroshnichenko2013} \\
    Sh 2$-$27 & $(102,14,45)$ & $17$ & \cite{vanleeuwen2007,harveysmith2011} \\
    Sh 2$-$205 & $(-852,524,-3.5)$ & $12$ & \cite{romero2008} \\
    S147 & $(-1199,-7,-34)$ & $32$ & \cite{sofue1980,kramer2003,chen2017}\\
    \end{tabular}
    \caption{Spatial parameters used to model the ISM features shown in Figure~\ref{fig:ism3d} and discussed in Section~\ref{sec:ism3d}. Central coordinates are shown in the heliocentric Galactic Cartesian reference frame with the Sun located at $(X,Y,Z) = (0,0,0)$. All features with a single radius are modeled as 3D spheroids, whereas multiple radii indicate an ellipsoid. The Local Bubble was modeled using a spherical harmonic expansion of its dust extinction boundary \citep{pelgrims2020}. The spatial extents of molecular clouds are not modeled and only their positions are shown in Figure~\ref{fig:ism3d}. Details are provided for each ISM feature in the main text of Appendix~\ref{app:ISM}.}
    \label{tab:ism_features}
\end{table*}

The ISM structures shown in Figure~\ref{fig:ism3d} are largely drawn from models in the literature based on multi-wavelength surveys. Here we describe the basic parameters of each modeled feature in Figure~\ref{fig:ism3d} and their basis in observations. The coordinates and literature references used for each ISM feature are shown in Table~\ref{tab:ism_features}.

\textbf{Local Bubble:} We adopt the \cite{pelgrims2020} model for the inner surface of the Bubble, which is based on comprehensive 3D dust extinction maps of the local ISM derived from Gaia and 2MASS data \citep{lallement2019}. \cite{pelgrims2020} define the inner surface of the Bubble as the first inflection point in the differential extinction as a function of distance from the Sun, and provide a spherical harmonic expansion of the corresponding 3D Bubble surface. Note that this model does not correspond to the peak dust extinction along any given LOS, and that the spherical harmonic decomposition can differ dramatically from the raw dust extinction data for some LOSs, particularly for lower multipole degrees. We use the Pelgrims model with a spherical harmonic multipole degree $l = 6$, which captures most of the large-scale variations across the Bubble's surface without giving too much weight to smaller scale features that are limited by noise and the angular resolution of the dust extinction maps. 
 
\textbf{GSH 238$+$00$+$09:} This superbubble was originally identified by \cite{heiles98} in IR, X-ray, and radio images and has since been observed in dust extinction maps \citep{lallement2014}. The superbubble is centered on a Galactic longitude of about $238^\circ$ and extends from about 0.2 to 1.2 kpc from the Sun, with a width in the Galactic plane of about 300 pc. Given that no detailed model, analogous to the Pelgrims model for the Local Bubble, exists for the surface of this superbubble, we model the superbubble as a 3D ellipsoid with the parameters given in Table~\ref{tab:ism_features} based on \cite{lallement2014}. However, we note that \cite{ocker2020} attributed a large DM deficit observed for J1024$-$0719 to this superbubble, which would imply a larger extent perpendicular the Galactic plane ($Z$) than inferred from dust extinction and as shown in the 3D interactive version of Figure~\ref{fig:ism3d}.

\textbf{Per-Tau Shell:} Using parsec-resolution dust maps \citep{leike2020} of the star-forming regions encompassing the Perseus and Taurus molecular clouds, \cite{bialy2021} identified an outer shell of dust containing an HI shell filled by concentrated H$\alpha$ emission. We use spheroidal model parameters provided by \cite{bialy2021} to include the Per-Tau shell in Figure~\ref{fig:ism3d}.

\textbf{Molecular Clouds:} The positions of molecular clouds are taken from \cite{zucker2020}, who provide a uniform catalog of distances to 60 of the major molecular clouds within 2 kpc of the Sun based on Gaia DR2 parallaxes and stellar photometric data from DECam. Figure~\ref{fig:ism3d} shows only the positions of these clouds; their spatial shapes and extents are not modeled.

\textbf{Sh 2$-$7:} This HII region is centered around a Galactic longitude and latitude of $(-9^\circ,24^\circ)$ and is energized by the star $\delta$ Scorpii at a distance of 136 pc \citep{miroshnichenko2013}. Using the H$\alpha$ map from \cite{finkbeiner2003} we estimate the diameter of the HII region to be about 28 pc and model the region as a simple spheroid located at the distance of $\delta$ Scorpii.

\textbf{Sh 2$-$27:} This HII region is centered around Galactic coordinates $(8^\circ,24^\circ)$ and is energized by the star $\zeta$ Ophiuchi at a distance of 112 pc \citep{vanleeuwen2007}. The projected diameter of the HII region in H$\alpha$ emission is about 34 pc \citep{harveysmith2011}. We model the region as a simple spheroid. Both Sh 2$-$7 and Sh 2$-$27 are part of the larger Ophiuchus cloud complex, and observations of molecular gas suggest the cloud complex as a whole has a central distance of about 140 pc with a depth of about 120 pc \citep{degeus1991}.

\textbf{Sh 2$-$205:} This HII region has Galactic coordinates $(148.4^\circ,-0.2^\circ)$ and is energized by the star HD 24431 at a distance of about 1 kpc \citep{romero2008}. We use the estimated spatial parameters provided by \cite{romero2008} to model the region as a simple spheroid, although \cite{romero2008} find that comparison of H$\alpha$, HI, and radio data for the region suggests it consists of three distinct substructures.

\textbf{S147:} This supernova remnant associated with PSR J0538$+$2817 is located around Galactic coordinates $(180^\circ,-1.6^\circ)$ \citep{kramer2003} with a projected radius in radio images of about 32 pc \citep{sofue1980,chen2017}. We use a spheroidal model with the same parameters assumed by \cite{yao2021} in their analysis of the scintillation produced by S147. 

\bsp	
\label{lastpage}
\end{document}